\documentclass[a4paper,12pt]{article}
\usepackage[a4paper,margin=1in]{geometry}
\usepackage{amsmath}
\usepackage{amssymb}
\usepackage{graphicx}
\usepackage{hyperref}
\usepackage{xcolor}
\usepackage[normalem]{ulem}
\usepackage[version=4]{mhchem}
\usepackage{algorithm}
\usepackage{algpseudocode}
\usepackage{enumitem}
\usepackage{float}
\usepackage{booktabs}
\usepackage{tabularx}
\usepackage{array}
\usepackage{caption}
\usepackage{colortbl}
\usepackage{authblk}
\usepackage{longtable}
\usepackage[sorting=none]{biblatex}

\title{Active Learning Discovery of High Temperature Oxidation Resistant Refractory Complex Concentrated Alloys}

\usepackage{authblk}
\title{Active Learning Discovery of High Temperature Oxidation Resistant Refractory Complex Concentrated Alloys}

\author[1,*]{Akhil Bejjipurapu}
\author[2]{Sharmila Karumuri}
\author[1]{Joseph C. Flanagan}
\author[1]{Victoria Tucker}
\author[2]{Ilias Bilionis}
\author[1]{Alejandro Strachan}
\author[1]{Kenneth H. Sandhage}
\author[1]{Michael S. Titus}

\affil[1]{School of Materials Engineering, Purdue University, West Lafayette, IN 47907, USA}
\affil[2]{School of Mechanical Engineering, Purdue University, West Lafayette, IN 47907, USA}

\affil[*]{Corresponding author: abejjipu@purdue.edu}

\newcommand{\SK}[1]{\textcolor{black}{#1}}

\begin{document}
\maketitle
\begin{abstract}
Refractory complex concentrated alloys (RCCAs) are of significant interest for advanced high-temperature applications, owing to their broad compositional range and potential for attractive mechanical properties and oxidation resistance. However, their compositional complexity poses significant challenges to conventional alloy discovery methodologies. In this study, an active learning framework is introduced that integrates Gaussian process regression with Bayesian global optimization to accelerate identification of oxidation-resistant RCCAs. Focusing on aluminum-containing quaternary systems, alloy and oxide descriptors were used to predict oxidation performance at 1000\textdegree{}C. Beginning with a dataset of 81 experimentally validated RCCAs, this framework was used to iteratively select alloy batches (five alloys per batch) with optimization based on a balance between exploration and exploitation to minimize associated experimental costs. After six iterations, two alloys were identified (nominal Al$_{30}$Mo$_5$Ti$_{15}$Cr$_{50}$ and Al$_{40}$Mo$_5$Ti$_{30}$Cr$_{25}$) that exhibited specific mass gains less than 1 mg/cm$^2$ at 1000\textdegree{}C in air. Both of these alloys formed adherent external $\alpha$-Al$_2$O$_3$ scales and exhibited parabolic oxidation kinetics consistent with diffusion-limited scale growth. Furthermore, our multiobjective analysis demonstrates that these alloys simultaneously achieve high specific hardness ($>0.12$ HV$_{0.5}$m$^3$/kg) and thermal expansion compatibility with thermal barrier coating systems, positioning them as promising bond coat candidates. This work underscores the efficacy of active learning in traversing complex compositional landscapes, and offers a scalable strategy for the development of advanced materials suitable for extreme environments.
\end{abstract}
\vspace{1em}
\newpage
\section*{Introduction}

The development of materials capable of withstanding extreme environments, particularly at elevated temperatures, remains a critical challenge for the aerospace, energy, and defense sectors. Refractory complex concentrated alloys (RCCAs), which  comprise a subclass of multi-principal element alloys that incorporate refractory metals such as Ti, Zr, Hf, V, Nb, Ta, Cr, Mo, and W, present promising alternatives to conventional alloys for high-temperature applications  \cite{Senkov2010RefractoryAlloys, Senkov_Miracle_Chaput_Couzinie_2018, Senkov2014MicrostructureAlloys}. Nevertheless, the potential use of RCCAs is often curtailed by insufficient environmental resistance. In particular, the inability to develop dense, slow-growing, adherent (protective) oxide scales under oxidative conditions compromises their high-temperature performance. One critical application area where oxidation resistance is paramount is in thermal barrier coating (TBC) systems used in gas turbines and aerospace engines. In such systems, the bond coat plays a dual role of promoting adherence of the ceramic topcoat and of forming a protective oxide barrier  \cite{Clarke_Oechsner_Padture_2012, Shahbazi2024HighReview}. MCrAlY (M = metal, Y = small elemental editions)-based bond coats are commonly used, but their limited compatibility with lightweight substrates, such as $\gamma$-TiAl, has prompted interest in alternative compositions  \cite{Pflumm2015OxidationReview, Tang2002DevelopmentAlloys}. RCCAs with tailored Al and Cr contents offer a promising new class of bond coat candidates with the potential to address both oxidation and substrate compatibility requirements.

In contrast to typical RCCAs, advanced nickel-based superalloys with as little as 8--9 at.\% Al effectively form protective $\alpha$-Al$_2$O$_3$ scales for long-term, high-temperature environmental resistance \cite{Lo2022ElementalAlloys}. Perhaps surprisingly, even with up to 20 at.\% Al, RCCAs frequently fail to achieve similar protective behavior. Earlier efforts to develop CrTaO$_4$-based oxide layers on RCCAs have only partially addressed oxidation resistance, as these scales are prone to internal oxidation,  with deleterious species such as nitrogen penetrating into the substrate and contributing to embrittlement  \cite{Gorr2020ATemperatures, Schellert2021TheTa-Mo-Cr-Ti-xAl, Lo2019AnOxide, Muller2019OnAlloys}. Hence, the identification of RCCAs capable of forming continuous, adherent, slow-growing, external $\alpha$-Al$_2$O$_3$scales for achieving high-temperature oxidation resistance remains a strong desire.

A major challenge with RCCAs lies in their vast composition space, making traditional methods like trial-and-error and DFT-based calculations time consuming and inefficient. Although CALPHAD-based modeling has been attempted  \cite{Bosi2023EmpiricalAlloys, Gorsse2018CurrentAlloys, Senkov2019CALPHAD-aidedNbTiZr}, it remains costly and time intensive. Consequently, leveraging advances in artificial intelligence and machine learning, particularly active learning methods, has gained traction. These methods utilize a combination of machine learning (ML) surrogates and Bayesian optimization to efficiently navigate the expansive composition space, so as to identify a limited number of promising alloys for testing~\cite{Hastings2025AcceleratedSpace, Khatamsaz2023BayesianApproach, kotthoff2021bayesianoptimizationmaterialsscience, Paramore2025Two-shotAlloys, JangActiveThermoelectrics, Tian2024Noise-AwareHeat, Thelen2025AcceleratingReaction}. This methodology can accelerate the discovery process and thereby enhance the efficiency and impact of materials research.

While machine learning (ML) has been successfully applied to predict properties such as yield strength \cite{Bhandari2021YieldLearning, Klimenko2021MachineSystem}, hardness \cite{Ye2023ImprovingData, Gao2023MachineAlloys}, and low-temperature corrosion resistance \cite{Zou2024MachineAlloys}, the application of ML for identifying RCCAs with high-temperature oxidation resistance has not been as extensive. Several studies have highlighted the capability of ML to identify particular oxidation-related characteristics, including mass change  \cite{Loli2022PredictingMethods, Duan2023DesignLearning, YunModellingNetwork}, oxide scale thickness \cite{BiancoPredictingApproach}, effective activation energy (Arrhenius behavior) \cite{WeiDiscoveringLearning, TaylorHighLearning}, and the associated parabolic rate constant ($k_p$) \cite{TaylorHighLearning,Bhattacharya2020PredictingLearning,Aghaeian2023PredictingModels}. For example, Kim \textit{et al.} \cite{KimRegressionNetwork} applied an artificial neural network (ANN) to forecast mass gain in a complex Ni-based alloy system comprising Ni-Co-Cr-Mo-W-Al-Ti-Ta-C-B, using proprietary high-throughput experimental datasets. Duan \textit{et al.} \cite{Duan2023DesignLearning} developed a model to predict mass change in Ni-based superalloys within the Ni-Co-Cr-Mo-Al-Ti-Nb-Hf-Zr alloy system. Bhattacharya \textit{et al.} \cite{Bhattacharya2020PredictingLearning} constructed a regression model to estimate $k_p$ values for Ti-based alloys, and Dewangan \textit{et al.} \cite{Dewangan2022ApplicationAlloys} utilized an ANN to predict mass gain in AlCrFeMnNiW$_x$ alloys (where $x = 0, 0.05, 0.1, 0.5$). 
\SK{Mishra \textit{et al.} \cite{Mishra2024MassSharing} developed an open-source computational workflow that systematically fits mass-gain data to multiple oxidation kinetics models and leverages Bayesian statistics for robust model selection}.
Despite these advancements, prior work has largely centered on optimizing oxidation resistance within restricted compositional domains or clarifying empirically observed trends between alloying elements and oxidation behavior. Notably, Gorsse \textit{et al.} \cite{Gorsse2025AdvancingResistance} recently employed Gradient Boosted Decision Trees (GBDT) to predict specific mass gain due to oxidation of refractory high-entropy alloys (RHEAs), and demonstrated significant accuracy while exploring an extensive compositional design space. However, most existing models do not incorporate oxide-specific thermodynamic features, and few studies have extended to broader design spaces or to iteratively-refined predictions through feedback from experiments.

\begin{figure}[H]
    \centering
    \includegraphics[width=\textwidth]{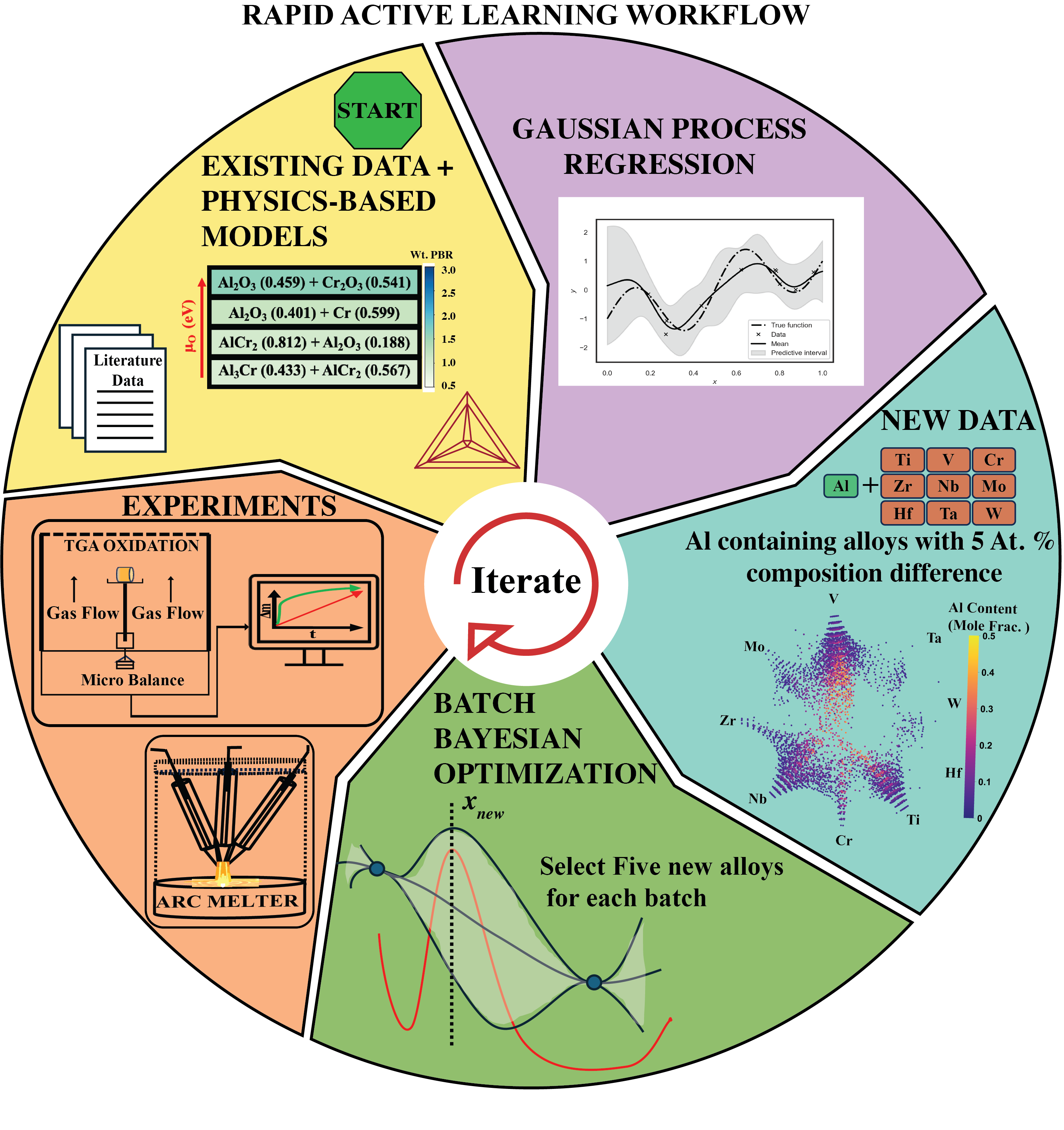}
    \caption{Schematic illustration of the rapid active learning workflow for discovering oxidation-resistant RCCAs, revealing the iterative batch process of combining existing data, Gaussian Process Regression, batch Bayesian optimization, experimental synthesis, and data augmentation for selecting five new alloys for each iteration or batch.}
    \label{fig:workflow}
\end{figure}

In this study, we introduce an innovative active learning framework that integrates Gaussian process regression (GPR) with \SK{Batch-}Bayesian global optimization (\SK{Batch-}BGO) to efficiently and rapidly discover oxidation-resistant RCCAs. This methodology was adapted from an approach previously developed by Karumuri \textit{et al.} \cite{Austin-Herna2025} to discover alloys with high hardness. The workflow, illustrated in Figure~\ref{fig:workflow}, synergistically combines alloy- and oxide-specific descriptors—including CALPHAD-derived phase stability and oxidation-related thermodynamic data—to predict oxidation resistance at 1000\textdegree{}C in air. The process involves defining the alloy design space, preparing the initial dataset, calculating descriptors, building a predictive GPR model, and optimizing the selection of alloys through BGO using Expected Improvement (EI) as the acquisition function. Starting with a dataset of 81 compositions obtained from the literature, we implemented this active learning loop to select and experimentally evaluate batches of five new alloys per iteration. Over six cycles, our model-guided investigation identified two quaternary alloys that exhibit slow, and decreasing, rates of oxidation resulting from the formation of dense, adherent, external $\alpha$-Al$_2$O$_3$ scales. This study demonstrates the efficacy of active learning in accelerating alloy discovery, while minimizing experimental workload. Our results provide a promising strategy for identifying advanced, oxidation-resistant RCCAs for future use under extreme service conditions.

\section*{Results}

\begin{figure}[H]
    \centering
    \includegraphics[width=\textwidth]{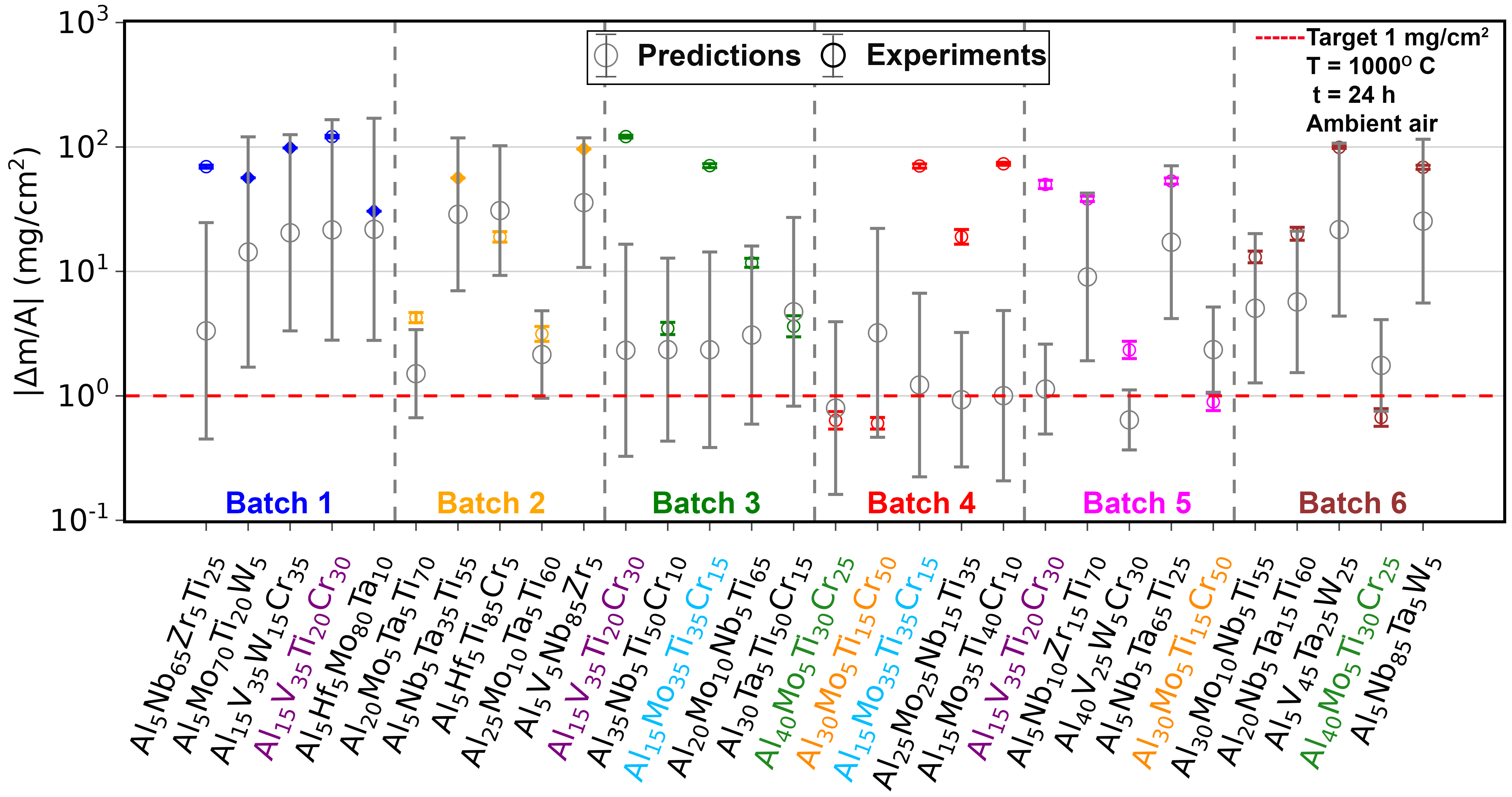}
    \caption{Predicted and experimentally-measured specific mass gain values for 24 h of oxidation at 1000\textdegree C in ambient air for 30 alloys evaluated across six batches. Solid circles represent experimental values with error bars, open circles show GPR predictions with error bars, and colors denote different batches (Batch 1: green, Batch 2: blue, Batch 3: cyan, Batch 4: red, Batch 5: magenta, Batch 6: brown). The red dashed line indicates the target threshold (1 mg/cm$^2$) for oxidation resistance. }
    \label{fig:pred_vs_exp}
\end{figure}

The predicted and experimentally measured specific mass gain values for 30 alloys obtained from our BGO approach, evaluated across six batches, are presented in Figure~\ref{fig:pred_vs_exp} (oxidation for 24 h at 1000\textdegree{}C in ambient air). Each data point represents the nominal (targeted) unique alloy composition, with the error bars ($\mu \pm \sigma$) indicating the uncertainty of the GPR model predictions.

The nominal (targeted) and measured compositions (from EDX analyses) for all 30 evaluated RCCAs, along with corresponding values of predicted and experimentally-measured specific mass gains, are provided in SI Table~2. The presence of some repeated alloy entries across multiple batches resulted from the iterative model update strategy, which involved the recalculation of descriptors using measured alloy compositions (rather than nominal values). This approach ensured more accurate representation of alloy chemistries in subsequent GPR training cycles, but also resulted in some nominal compositions being re-selected when the measured composition deviated from the nominal composition, and when the nominal composition can yield high expected improvement under an updated model.

The initial exploratory alloys in Batch 1 exhibited poor oxidation resistance, with specific mass gains exceeding 30 mg/cm$^2$. Notable examples included the nominal compositions, Al$_5$Nb$_{65}$Zr$_5$Ti$_{25}$ and Al$_{15}$V$_{35}$Ti$_{20}$Cr$_{30}$, for which specific mass gains were measured to be in excess of 90 mg/cm$^2$. This batch established a necessary baseline for the active learning loop by sampling a wide range of alloy compositions. 

For Batches 2 and 3, modest improvements in oxidation resistance (to below 5 mg/cm$^2$) were observed for several alloys (including the nominal compositions: Al$_{20}$Mo$_5$Ta$_5$Ti$_{70}$, Al$_{25}$Mo$_{10}$Ta$_5$Ti$_{60}$, Al$_{35}$Nb$_5$Ti$_{50}$Cr$_{10}$ and Al$_{30}$Ta$_5$Ti$_{50}$Cr$_{15}$) although the specific mass gains for all of these alloys exceeded the targeted threshold value of 1 mg/cm$^2$. For Batches 4 through 6, the model began to repeatedly target particularly promising alloys. The nominal compositions Al$_{30}$Mo$_5$Ti$_{15}$Cr$_{50}$ and Al$_{40}$Mo$_5$Ti$_{30}$Cr$_{25}$, were selected in Batch 4 (red). One of these alloys, Al$_{30}$Mo$_5$Ti$_{15}$Cr$_{50}$, was re-selected in Batch 5 (magenta) and the other alloy, Al$_{40}$Mo$_5$Ti$_{30}$Cr$_{25}$, was re-selected in Batch 6 (brown). These compositions emerged as the best-performing alloys, and achieved reproducible mass gains below the 1 mg/cm$^2$ threshold. The measured specific mass gain values for these two alloys also aligned closely with model predictions, with the narrowing error bars in later batches indicating increased model confidence in the performance of these compositions.

\begin{figure}[H]
    \centering
    \includegraphics[width=\textwidth]{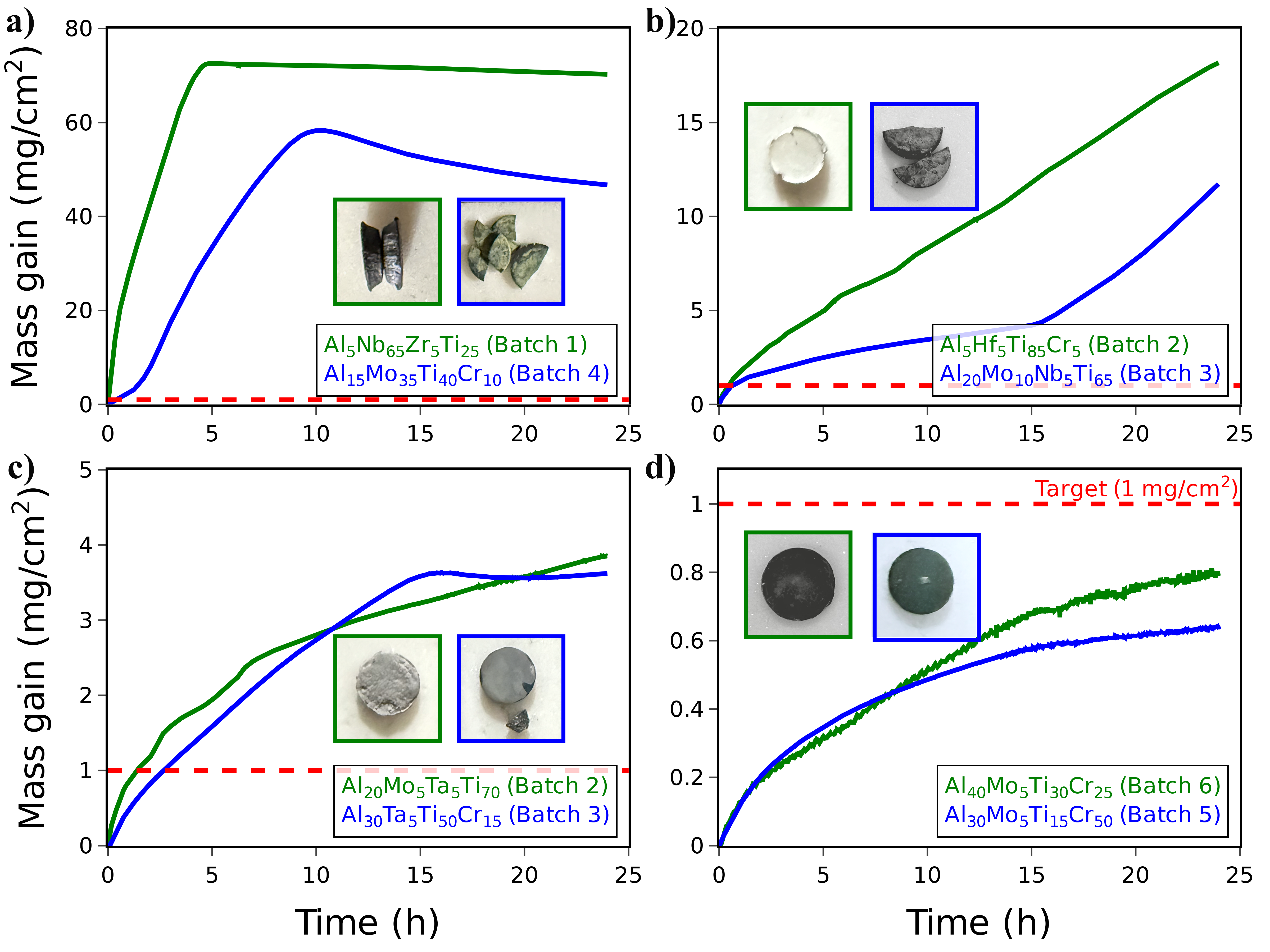}
    \caption{Continuous specific mass gain curves vs. time during oxidation in ambient air at 1000\textdegree C over 24 h (from TGA experiments), and post-oxidation specimen photographs, of selected alloys (nominal alloy compositions indicated). (a) Al$_5$Nb$_{65}$Zr$_5$Ti$_{25}$ (Batch 1, green) and Al$_{15}$Mo$_{35}$Ti$_{40}$Cr$_{10}$ (Batch 4, blue). (b) Al$_5$Hf$_5$Ti$_{85}$Cr$_5$ (Batch 2, green) and Al$_{20}$Mo$_{10}$Nb$_5$Ti$_{65}$ (Batch 3, blue). (c) Al$_{20}$Mo$_5$Ta$_5$Ti$_{70}$ (Batch 2, green) and Al$_{30}$Ta$_5$Ti$_{50}$Cr$_{15}$ (Batch 3, blue). (d) Al$_{30}$Mo$_5$Ti$_{15}$Cr$_{50}$ (Batch 5, blue) and Al$_{40}$Mo$_5$Ti$_{30}$Cr$_{25}$ (Batch 6, green).}
    \label{fig:tga_curves}
\end{figure}

Plots of continuous specific mass gain vs. time obtained during the oxidation of representative alloys in ambient air at 1000\textdegree C for 24 h from TGA experiments are provided in Figures~\ref{fig:tga_curves} (a--d). Post-exposure photographs of the alloy samples associated with these plots are also provided. These examples reveal the transition from high mass gain to low mass gain compositions (or from non-protective oxide formation to more protective, adherent oxide scale formation). Figure~\ref{fig:tga_curves}a provides data for the nominal alloy compositions: Al$_5$Nb$_{65}$Zr$_5$Ti$_{25}$ (Batch 1) and Al$_{15}$Mo$_5$Ti$_{10}$Cr$_{30}$ (Batch 4). The former alloy (green curve) exhibited complete oxidation with a mass gain plateau near 70 mg/cm$^2$, and this oxidized specimen was heavily distorted relative to the starting disk shape. The latter alloy (blue curve) reached a maximum mass gain near 60 mg/cm$^2$, and then exhibited a mass loss. This specimen had broken into multiple pieces during the oxidation process. 

Continuing our comparative analysis, Figure~\ref{fig:tga_curves}b shows Al$_5$Hf$_5$Ti$_{85}$Cr$_5$ (Batch 2) and Al$_{20}$Mo$_{10}$Nb$_5$Ti$_{65}$ (Batch 3).  After a few hours, the Al$_5$Hf$_5$Ti$_{85}$Cr$_5$ alloy sample (green curve) exhibited a nearly constant rate of oxidation, with a final specific mass gain of about 19 mg/cm$^2$. This specimen retained a general disk shape, but with raised edges, and turned white in color. For the Al$_{20}$Mo$_{10}$Nb$_5$Ti$_{65}$ alloy (blue curve), the rate of oxidation initially slowed with time, but then transitioned to a higher oxidation rate with a final specific mass gain of about 12 mg/cm$^2$. This specimen fractured into two pieces after completion of the oxidation test. Both of the alloys shown in Figure~\ref{fig:tga_curves}c, Al$_{20}$Mo$_5$Ta$_5$Ti$_{70}$ (Batch 2) and Al$_{30}$Ta$_5$Ti$_{50}$Cr$_{15}$ (Batch 3), exhibited more modest specific mass gain values $<$5 mg/cm$^2$ and retained the starting disk shapes. The latter alloy possessed a smoother surface after the oxidation test than did the former alloy. Figure~\ref{fig:tga_curves}d reveals the oxidation behavior of the most oxidation-resistant alloys, Al$_{30}$Mo$_5$Ti$_{15}$Cr$_{50}$ (Batch 5) and Al$_{40}$Mo$_5$Ti$_{30}$Cr$_{25}$ (Batch 6). For both alloys, the rate of oxidation slowed with time, and the final specific mass gains were below 1 mg/cm$^2$. After the oxidation test, these specimens exhibited excellent retention of the starting disk shapes with relatively smooth macrocrack-free surfaces.

\begin{figure}[H]
    \centering
    \includegraphics[width=\textwidth]{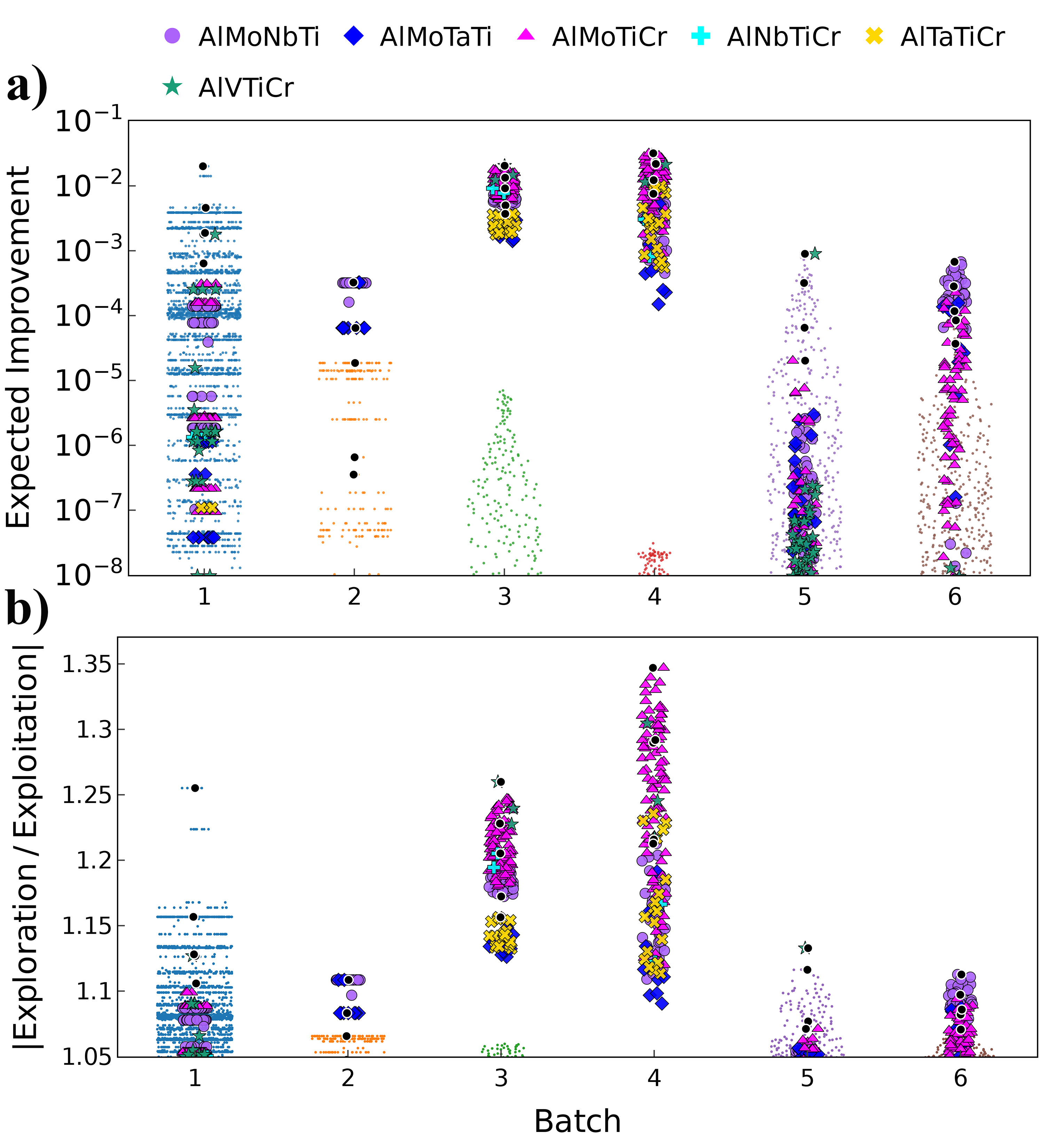}
    \caption{(a) Expected Improvement (EI) values for design space alloys across the six active learning batches, with each dot representing an alloy candidate, with high-EI candidates in Batches 3 and 4 highlighted using distinct symbols, and with black circles indicating selected alloys. (b) The Exploration-to-exploitation ratio ($|$exploration/exploitation$|$) across the six active learning batches.}
    \label{fig:ei_exploration}
\end{figure}

To obtain insight into the BGO process, the EI values of the examined alloys for design space alloys across all six active learning batches are presented in Figure~\ref{fig:ei_exploration}a. Each dot represents an alloy candidate, with families having high EI values ($>$10$^{-4}$) in Batches 3 and 4 highlighted using distinct symbols to track their progression. The absolute value of the exploration-to-exploitation ratio ($|\text{exploration}/\text{exploitation}|$) as a function of the batch number is provided in Figure~\ref{fig:ei_exploration}b. This total range of values for this ratio increased from Batches 2 to 4, and then decreased from Batch 4 to Batches 5 and 6, which was consistent with a transition from exploratory to exploitative behavior.

To further clarify this transition, plots of the exploration vs. exploitation values for every candidate alloy in each for the six batches are presented in SI Figure 1. The inset plots in Batches 2, 5, and 6 highlight low-exploration, high-exploitation candidates that were difficult to distinguish on the main axes due to scale differences.
In Batch 1, the model operated with high uncertainty with diverse compositions from various alloy families. The EI values for Batch 1 in Figure~\ref{fig:ei_exploration}a were broadly distributed, which was consistent with an exploration-dominant strategy. However, this exhaustive exploration yielded poor oxidation outcomes, as evidenced in Figure~\ref{fig:pred_vs_exp}, where most compositions exhibited high mass gains. The overall range of EI values for Batch 2 decreased as the first round of data was incorporated into the model and the predictions were refined. By Batch3, the overall range of EI values covered increased again, but distinct clusters of high-EI candidates emerged. This strategic shift from broad exploration to more targeted exploitation was driven by increased uncertainty due to aleatoric variability in the Batch 2 experimental results, as shown in Figure~\ref{fig:pred_vs_exp}. These elevated EI clusters indicated the identification of promising yet uncertain compositional regions.

In Batch 4, a cluster of high EI candidates that comprised six alloy families appeared, which indicated that higher-potential candidates began to be prioritized along with broad exploration. This critical round resulted in the identification of compositions that became high-confidence selections in Batches 5 and 6. As more data was incorporated into the model, the overall range of EI values decreased in Batches 5 and 6 relative to Batch 4. The data in Figure~\ref{fig:pred_vs_exp} and SI Figure~1 reflected this convergence, with several compositions from Batches 4--6 exhibiting significantly lower mass gains and clustering toward an exploitative regime. Early batches prioritized broad exploration to reduce uncertainty, while later batches focused on exploiting high-EI regions (identified in Batches 3 and 4), leading to the selection of top-performing compositions

\begin{figure}[H]
    \centering
    \includegraphics[width=\textwidth]{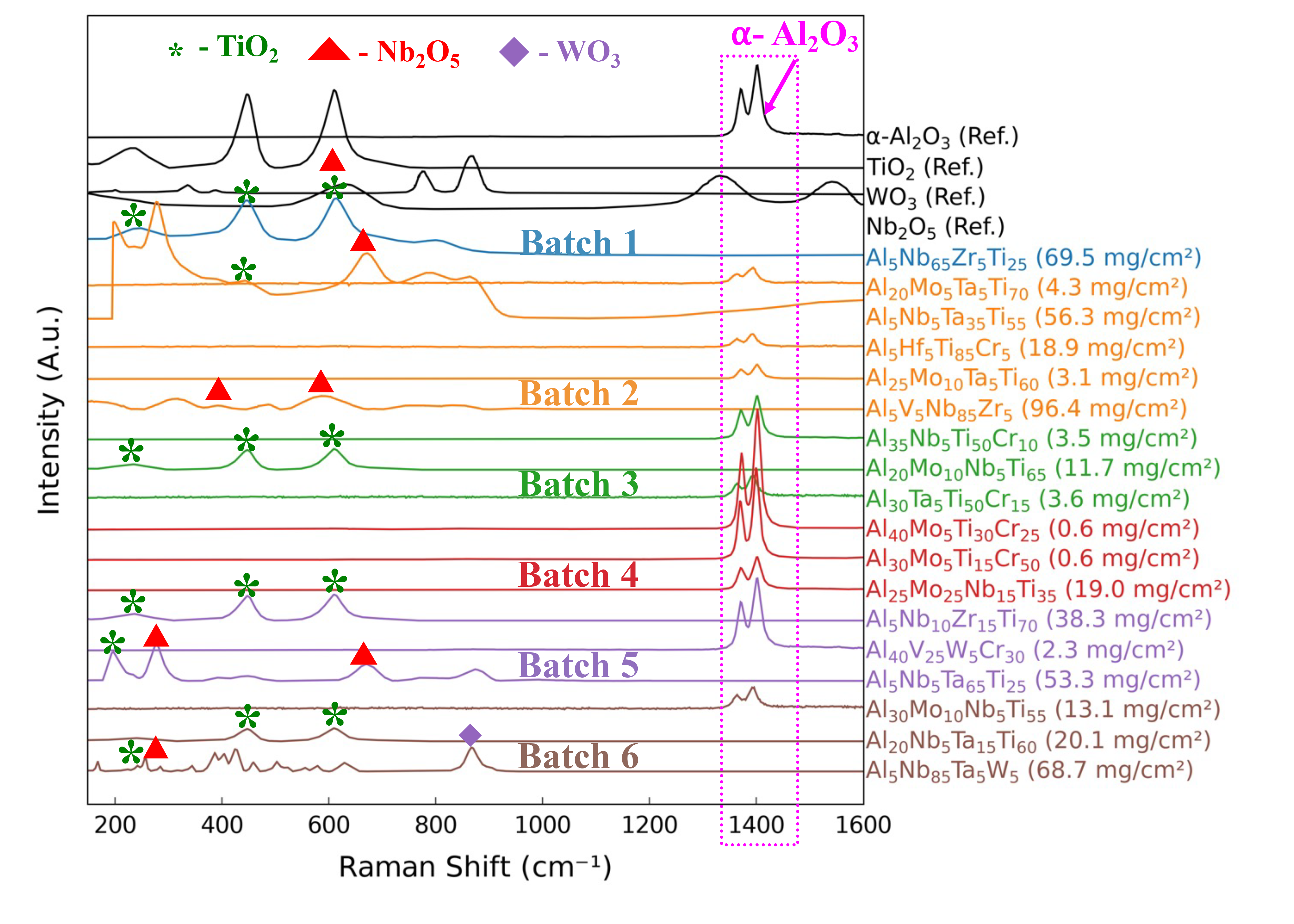}
    \caption{Raman spectra of representative alloys revealing peak signatures for  $\alpha$-Al$_2$O$_3$ (a characteristic fluorescent doublet indicated in the dotted rectangle), rutile TiO$_2$, Nb$_2$O$_5$, and WO$_3$.}
    \label{fig:raman_spectra}
\end{figure}

Raman spectra obtained from the surfaces of representative oxidized alloy specimens are provided in Figure~\ref{fig:raman_spectra}. Notably, ten alloys exhibited the fluorescent doublet peaks (near the Raman shift of 1400 cm$^{-1}$), that are characteristic of $\alpha$-Al$_2$O$_3$. These same $\alpha$-Al$_2$O$_3$ forming alloys exhibited relatively low specific mass gains (down to 0.60 mg/cm$^2$) after oxidation for 24 h at 1000\textdegree C in air. Alloys for which oxides other than $\alpha$-Al$_2$O$_3$ were detected (e.g., rutile \ce{TiO2}, \ce{Nb2O5}, \ce{WO3}) tended to exhibit relatively large specific mass gains (up to 96 mg/cm$^2$) after such oxidation. The correlation between the oxidation mass gain and the type of detected oxide was further examined by plotting specific mass gain values against the combined intensities of the two Raman peaks for \ce{$\alpha$-Al2O3} or of the three peaks for rutile \ce{TiO2} in SI Figure 2. Inverse correlations were observed for \ce{$\alpha$-Al2O3} and rutile \ce{TiO2} in this figure; that is, other spectra revealed peaks consistent with rutile \ce{TiO2}, frequently present but not associated with reduced mass gain.

\begin{figure}[H]
    \centering
    \includegraphics[width=0.5\textwidth]{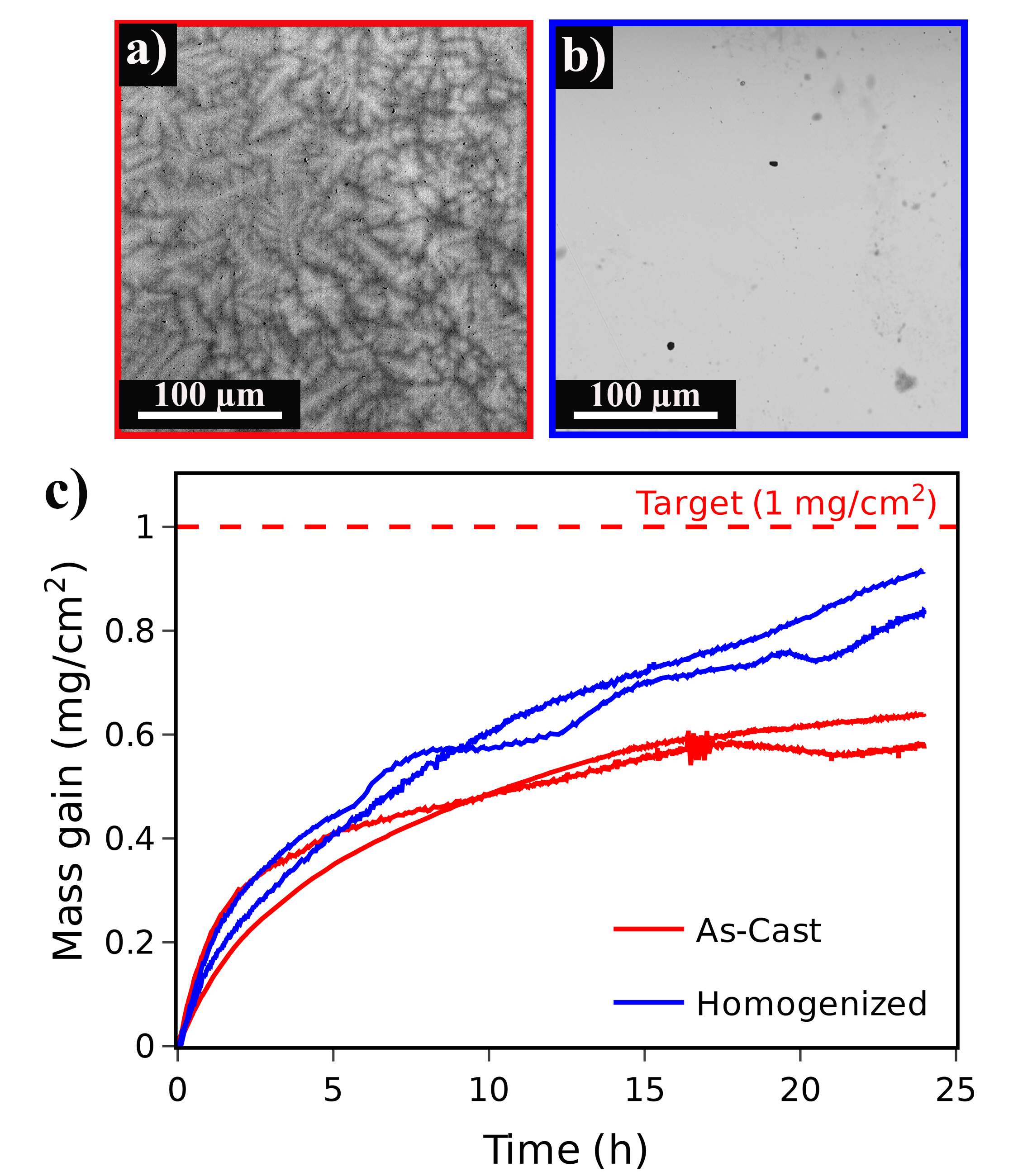}
    \caption{Microstructural analysis of the nominal Batch 5 alloy, \ce{Al30Mo5Ti15Cr50}, in the as-cast and homogenized conditions. Backscattered electron (BSE) images of polished cross-sections of: (a) the as-cast alloy and (b) the homogenized alloy (after vacuum heat treatment for 10h at 1300\textdegree C). (c) Specific mass gain curves (TGA) at 1000\textdegree C for the as-cast and homogenized \ce{Al30Mo5Ti15Cr50} alloy.}
    \label{fig:microstructure}
\end{figure}

Cross-sectional SEM analyses (SI Figures ~3 and 4) were then used to characterize the $\alpha$-\ce{Al2O3}-rich scales formed on the two alloys exhibiting the lowest specific mass gain values: Al$_{40}$Mo$_5$Ti$_{30}$Cr$_{25}$ and Al$_{30}$Mo$_5$Ti$_{15}$Cr$_{50}$. After exposure to air for 24 h at 1000\textdegree C, both  of these alloys developed dense, continuous external oxide scales that were rich in aluminum and oxygen, with no apparent microcracking or delamination of these scales.
Further analyses were then conducted on the $\alpha$-\ce{Al2O3}-forming alloy of nominal composition, \ce{Al30Mo5Ti15Cr50}, which exhibited the lowest specific mass gain after 24 h at 1000\textdegree C in air (down to 0.6 mg/cm$^2$). In the as-cast state, this alloy possessed a dendritic microstructure, as seen in the backscattered electron (BSE) image of a polished cross-section in Figure~\ref{fig:microstructure}a. Exposure of this alloy to a vacuum heat treatment for 10 h at 1300\textdegree C resulted in a notably more chemically-homogeneous microstructure, as revealed by the BSE image of polished cross-section in Figure~\ref{fig:microstructure}b. The influence of such chemical homogenization on the oxidation kinetics of this alloy was then examined (Figure~\ref{fig:microstructure}c). While the specific mass gains of both as-cast and vacuum-annealed versions of this alloy fell below the target value of 1 mg/cm$^2$, the as-cast specimens exhibited slightly lower specific gains than the vacuum-annealed specimens after about 5 h exposure to air at 1300\textdegree C. The chemical inhomogeneity of the dendritic (as-cast) version of this alloy did not appear to degrade oxidation resistance of the alloy. 

\begin{figure}[H]
    \centering
    \includegraphics[width=\textwidth]{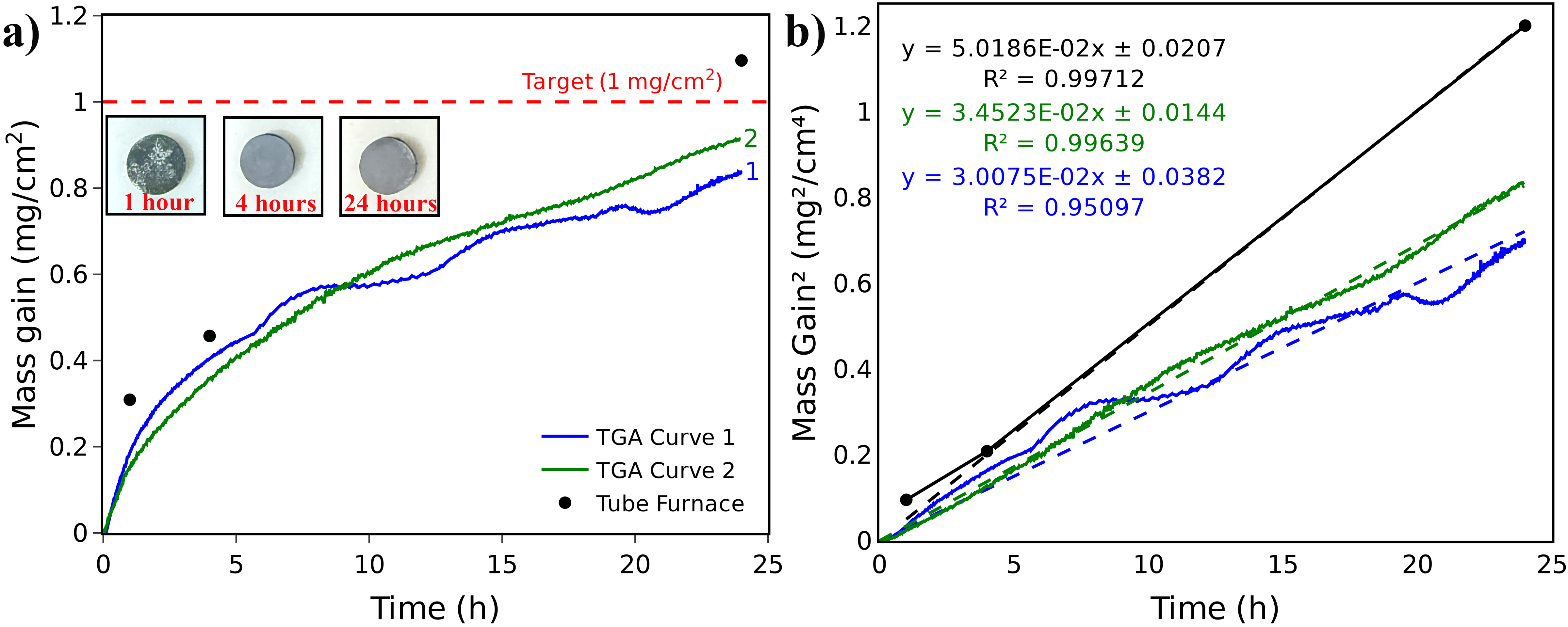}
    \caption{Oxidation kinetics obtained from interrupted and continuous oxidation experiments conducted on the vacuum-annealed (homogenized) Al$_{30}$Mo$_5$Ti$_{15}$Cr$_{50}$ alloy at 1000\textdegree C. (a) Specific mass gain data at 1, 4, and 24 h from interrupted (tube furnace) tests along with data from continuous TGA measurements, with optical images of specimen surfaces shown as insets. (b) Square of the specific mass gain values versus time.}
    \label{fig:interrupted_oxidation}
\end{figure}

Interrupted oxidation experiments were then conducted to examine the evolution with time of the scale that formed on the nominal \ce{Al30Mo5Ti15Cr50} alloy. Specific mass gain values are shown as individual data points in Figure~\ref{fig:interrupted_oxidation}a for vacuum -annealed \ce{Al30Mo5ti15Cr50} alloy specimens that had been rapidly inserted into the isothermal zone of a pre-heated tube furnace at 1000\textdegree C and then removed after intervals of 1, 4, and 24 h. The continuous specific mass gain values of such vacuum-annealed \ce{Al30Mo5Ti15Cr50} alloy specimens obtained at 1000\textdegree C from TGA experiments are also shown in Figure ~\ref{fig:interrupted_oxidation}a. The mass gain values from the interrupted oxidation experiments were in good agreement with the values obtained from the TGA experiments. Optical images of the surfaces of the specimens after 1, 4, and 24 h of oxidation (inset images in Figure ~\ref{fig:interrupted_oxidation}a) revealed a progressive change in color and color uniformity of the oxide scale between 1 h and 4-24 h. Plots the square of the specific mass gain against time shown in Figure ~\ref{fig:interrupted_oxidation}b reveal good fits to parabolic oxidation behavior (R$^2$  = 0.996 and 0.951), which were consistent with the diffusion-limited oxide growth of dense, adherent oxide scales.

\begin{figure}[H]
    \centering
    \includegraphics[width=\textwidth]{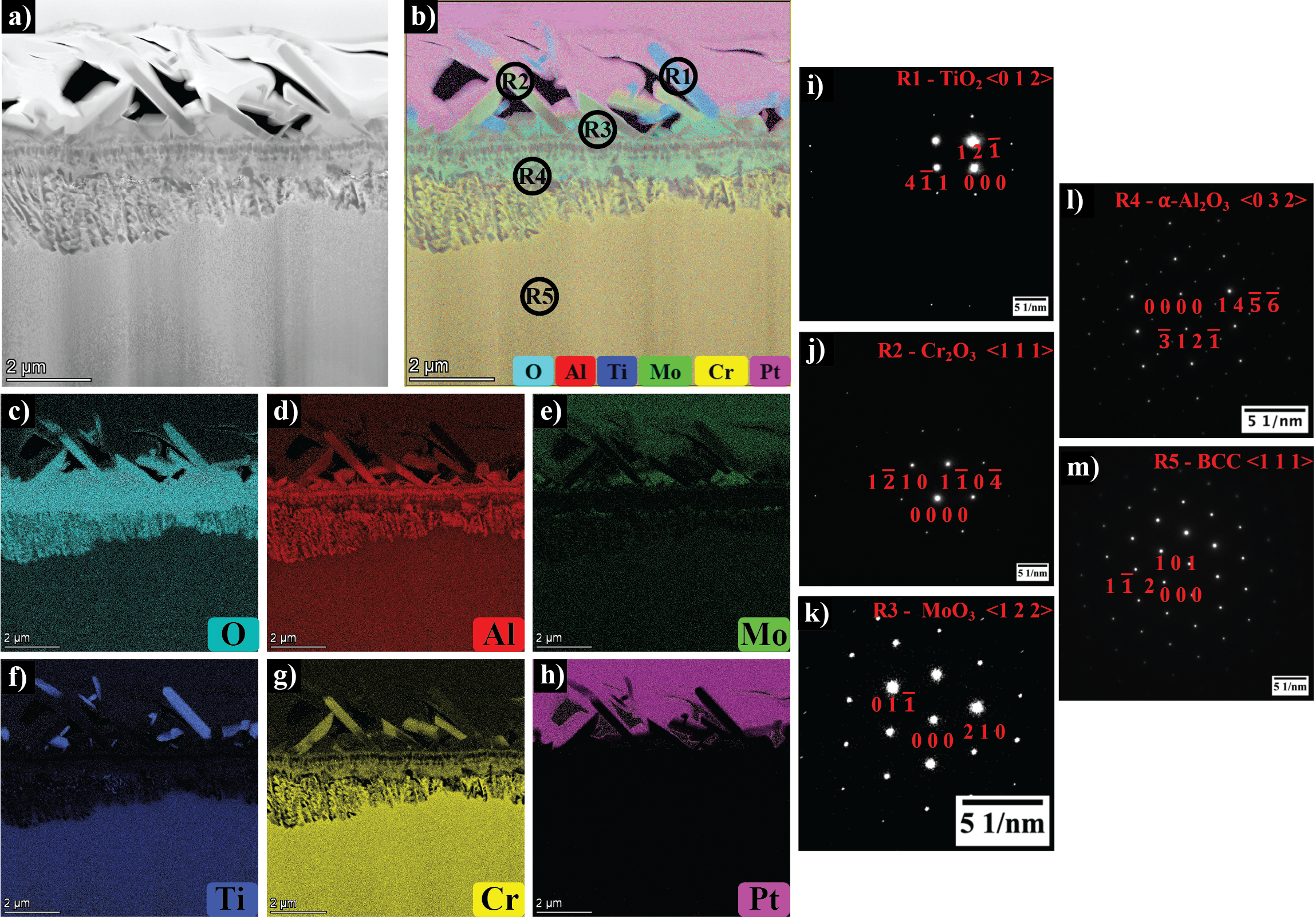}
    \caption{STEM-EDX elemental maps and SAED patterns obtained from an ion-milled cross-section of the oxide scale formed on the  homogenized nominal Al$_{30}$Mo$_5$Ti$_{15}$Cr$_{50}$ after exposure for 1 h to air at 1000\textdegree C.}
    \label{fig:stem_1hr}
\end{figure}

\begin{figure}[H]
    \centering
    \includegraphics[width=\textwidth]{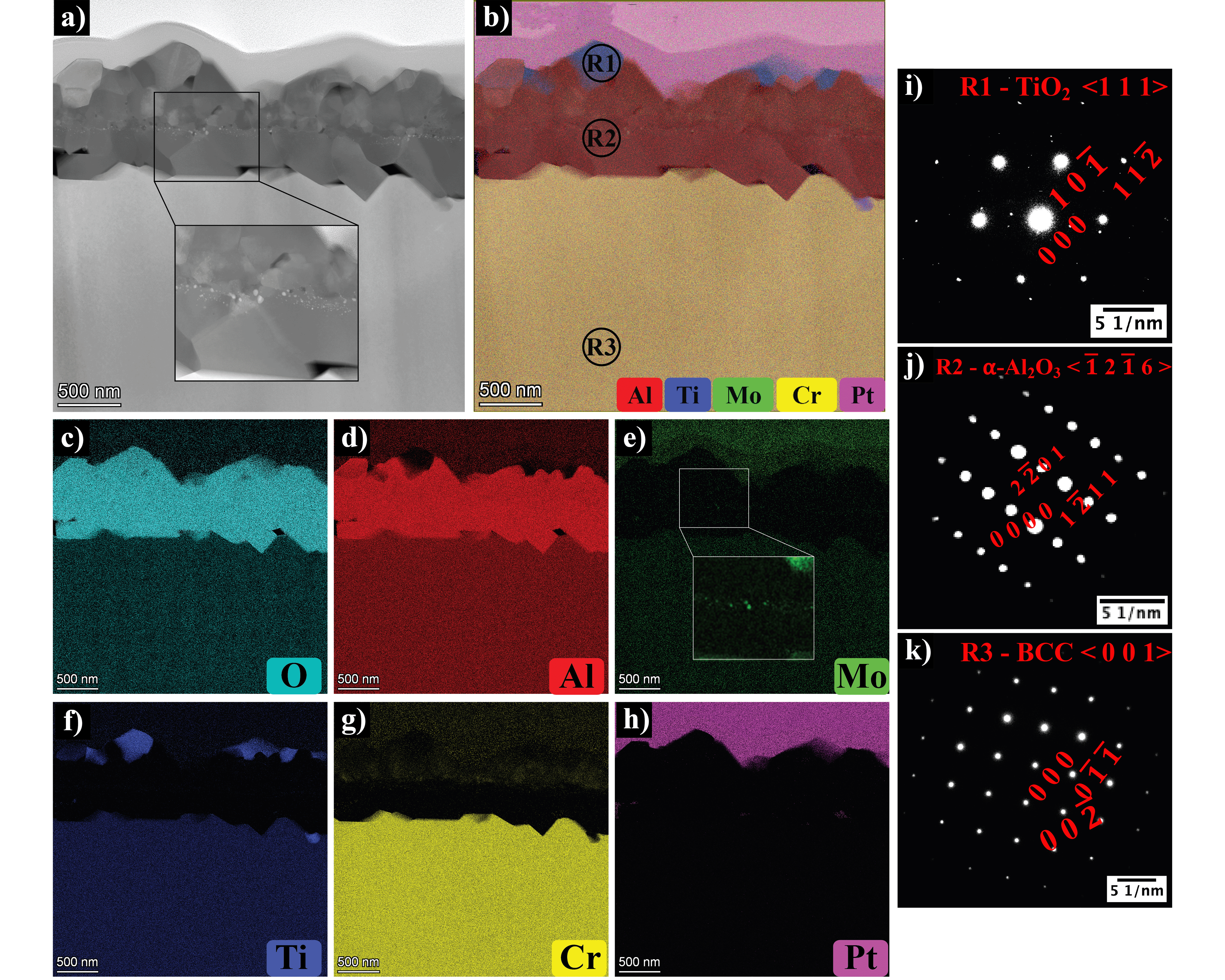}
    \caption{STEM-EDX elemental maps and SAED patterns obtained from an ion-milled cross-section of the oxide scale formed on the  homogenized nominal Al$_{30}$Mo$_5$Ti$_{15}$Cr$_{50}$ after exposure for 24 h to air at 1000\textdegree C.}
    \label{fig:stem_24hr}
\end{figure}

Scanning transmission electron microscopic (STEM) analyses, along with EDX and selected area electron diffraction (SAED) analyses, were then used to characterize ion-milled cross-sections of the oxide scales formed on the vacuum-annealed \ce{Al30Mo5Ti15Cr50} alloy after oxidation for 1 h and for 24 h at 1000\textdegree C. As revealed in Figure ~\ref{fig:stem_1hr}, a thin, multi-phase oxide scale had formed after just 1 h of oxidation at 1000\textdegree C. The elemental maps from EDX analyses indicated that relatively large, faceted Ti-O-rich and Cr-O-rich crystals had formed at the external scale surface, along with a smaller interspersed Mo-O-rich phase. SAED analyses indicated that the relatively high aspect crystals comprised rutile \ce{TiO2} and \ce{Cr2O3}, with the interspersed Mo-O-rich phase consisting of \ce{MoO3}. Al-O-rich particles were also present at the external surface. An underlying thin Al-O-rich layer was detected near the external surface, with internal Al-O-rich particles located below this thin layer. SAED analyses indicated that the Al-O-rich particles comprised $\alpha$-\ce{Al2O3}, and that the underlying homogenized metallic alloy possessed a BCC structure. After 24 h of oxidation at 1000\textdegree C (Figure ~\ref{fig:stem_24hr}), elemental maps from EDX analyses indicated that a continuous Al-O-rich external scale had formed, with sparse Ti-O-rich particles retained on the outside surface of this scale. Mo and Cr remained predominantly within the underlying alloy matrix. SAED patterns confirmed the presence of rutile \ce{TiO2} particles on the dense, external $\alpha$-\ce{Al2O3} scale, along with an underlying BCC alloy. The formation of such a continuous, dense, external $\alpha$-\ce{Al2O3} layer after 24 h in air at 1000\textdegree C was consistent with the small specific mass gain observed after such exposure for the nominal \ce{Al30Mo5Ti15Cr50} alloy, and underscored the importance of developing such an alumina scale, even after early-stage formation of other oxides (such as \ce{TiO2, Cr2O3, MoO3}), for long-term oxidation resistance.

\section*{Discussion}
\begin{figure}[H]
\centering
\includegraphics[width=\textwidth]{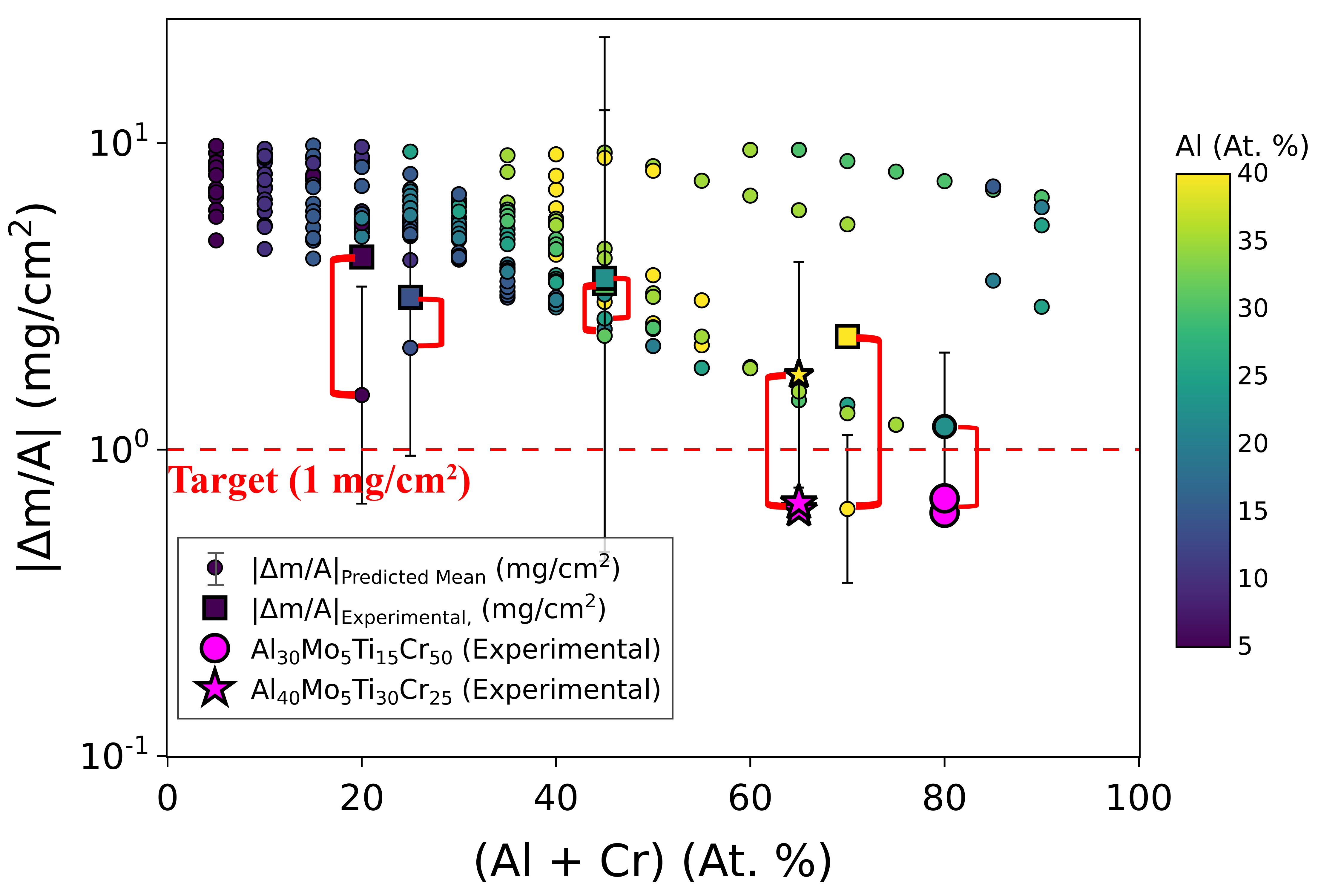}
\caption{Predicted and experimental oxidation mass gain versus nominal (Al + Cr) atomic percent for selected RCCA alloys. The color bar indicates Al content, with circles representing alloys with predicted mass gain less than 10 mg/cm$^2$ and squares representing alloys tested in our active learning batches with measured mass gain between 0 and 10 mg/cm$^2$. Red dashed line marks the 1 mg/cm$^2$ target. Highlighted nominal compositions Al$_{30}$Mo$_5$Ti$_{15}$Cr$_{50}$ (magenta circle) and Al$_{40}$Mo$_5$Ti$_{30}$Cr$_{25}$ (magenta star) achieve optimal performance, with their measured values shown as filled magenta symbols. For alloys with both predicted and experimental data, red brackets connect the predicted mean to their corresponding experimental values. Predicted standard deviations (±$\sigma$) are shown only for experimentally validated alloys to enhance visual clarity, as error bars for all predictions would extensively overlap due to high data density.}
\label{fig:alcr_plot}
\end{figure}

The concentrations of chromium and aluminum in refractory-metal-based alloys have been reported to be critical for forming continuous, slow-growing, protective scales—particularly comprising \ce{Cr2O3} and/or \ce{Al2O3} that are needed for achieving high-temperature oxidation resistance  \cite{Frohlich2010Ti-Al-Cr-TiAl, Brady1997TheOxygen, Fox-Rabinovich2012HierarchicalBehavior, Yan2009EvaluationAlloy, Brady2000AlloyFormation, Frohlich2008Ti-Al-Cr-XAluminides}. The correlation between predicted specific mass gain values and the combined Al and Cr contents of the nominal alloys of the present work is shown in Figure ~\ref{fig:alcr_plot}. Alloys with nominal total Al and Cr contents $>$65 at.\% tended to achieve specific mass gains below the 1 mg/cm$^2$ target (SI Table 2), and exhibited strong Raman peaks for $\alpha$-\ce{Al2O3} (Figure ~\ref{fig:raman_spectra}). Indeed, the top-performing nominal alloys, \ce{Al30Mo5Ti15Cr50} (Al+Cr = 80 at.\%) and \ce{Al40Mo5Ti30Cr25} (Al+Cr = 65 at.\%), exhibited specific mass gain values $<$0.90 mg/cm$^2$. It is worth noting that these two alloys formed protective $\alpha$-\ce{Al2O3} layers upon exposure to air for 24 h at 1000\textdegree C, as confirmed by Raman spectroscopy and STEM-EDX analyses (Figures ~\ref{fig:raman_spectra}, ~\ref{fig:stem_1hr}, ~\ref{fig:stem_24hr}), while possessing nominal Al contents of only 30–40 at.\% (EDX-measured Al contents of only 26.2-42.2 at.\%). A reduction in Al content can be important for reducing alloy brittleness associated with intermetallic formation. This result aligns with prior findings that Cr additions above 8–10 at.\% can reduce the Al content of Ti-Al-Cr alloys required to promote the formation of a continuous, protective $\alpha$-\ce{Al2O3} layer \cite{Brady1997TheOxygen, Fox-Rabinovich2006OxidationApplications, Yan2009EvaluationAlloy}

\begin{figure}[H]
\centering
\includegraphics[width=\textwidth]{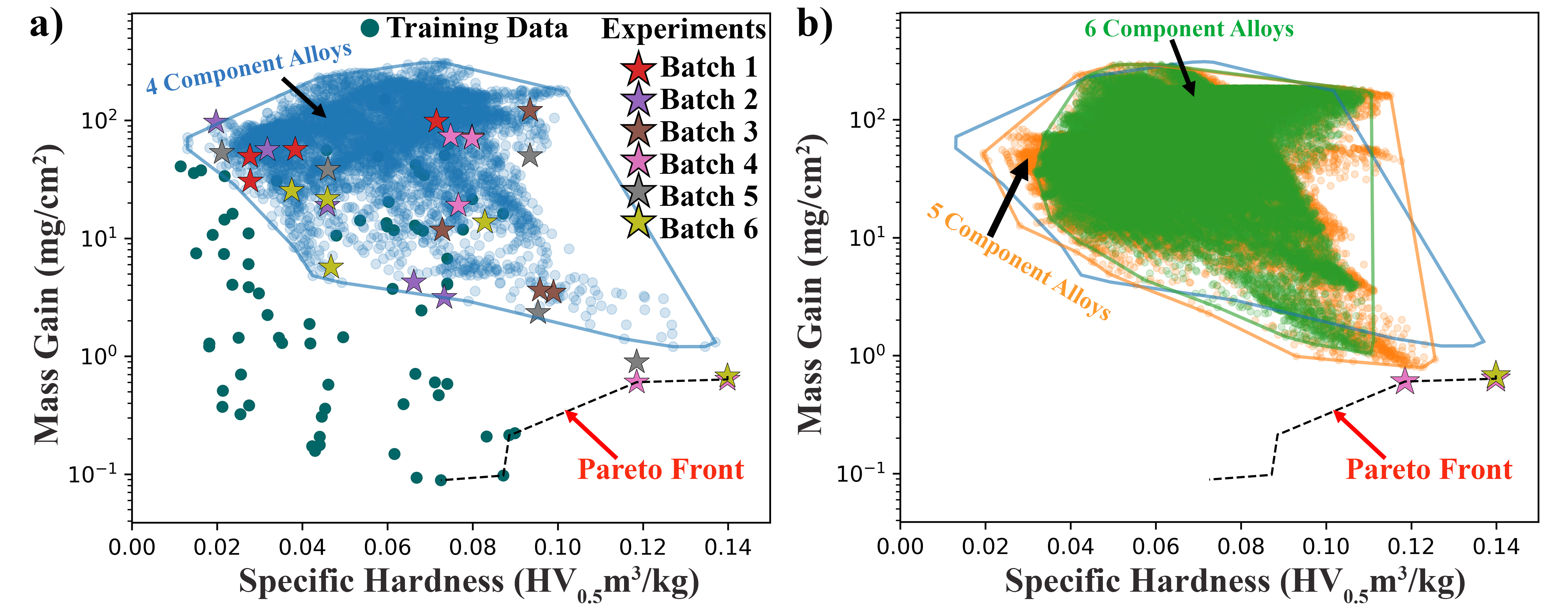}
\caption{Pareto front analysis for oxidation-induced specific mass gain and for specific hardness: (a) 4-component alloys with experimental batch results overlaid; (b) model-predicted properties for 5- and 6-component RCCAs. Black dashed lines trace the Pareto front.}
\label{fig:pareto_plots}
\end{figure}

To evaluate the trade-off between oxidation resistance and mechanical behavior, a two-objective Pareto front analysis was conducted for 4-component RCCAs. A plot of the predicted oxidation mass gain against specific hardness is provided in (Figure~\ref{fig:pareto_plots}a). The Pareto front (black dashed line) provides the composition of alloys that simultaneously achieved low specific mass gain and high specific hardness, which can be highly-desired attributes for structural materials in extreme environments. The training data and active learning-selected alloys (colored stars) demonstrate how this complex design space was navigated using the method of this work. Notably, the nominal alloy compositions Al$_{30}$Mo$_5$Ti$_{15}$Cr$_{50}$ and Al$_{40}$Mo$_5$Ti$_{30}$Cr$_{25}$ exhibited low measured specific mass gains below 1~mg/cm$^2$ and high measured specific hardness values (in excess of 0.12 HV$_{0.5}$m$^3$/kg and fell on the Pareto front, which validated the predictive capability of the Gaussian Process Regression (GPR) model in this work. This result demonstrates the ability of this approach for identifying compositions that balance often-competing chemical and mechanical properties, which is a key challenge for refractory alloys, including Ti-Al-Cr compositions for which high Al content can enhance oxidation resistance but at the expense of reduced ductility.

To further asses higher-order composition space, the design space was expanded to include 5- and 6-component RCCAs. This decision was motivated by the observation that, after Batch 4, the active learning loop began repeatedly selecting two similar quaternary compositions, along with a noticeable reduction in the range values of the exploration-to-exploitation ratio. 
Using the same descriptor framework and trained GPR model, the oxidation-induced specific mass gain and specific hardness values for the extended set of 5- and 6-component RCCAs were predicted. As shown in Figure~\ref{fig:pareto_plots}b, a shift occurred in the property landscape. The higher-order compositions clustered away from the 4-component Pareto front, exhibiting either increased predicted mass gains or decreased hardness values.

To validate these model predictions, two 5-component alloys with the lowest values of predicted oxidation-induced specific mass gain were selected for oxidation tests using the TGA: Al$_{30}$Mo$_{5}$V$_{5}$Ti$_{40}$Cr$_{20}$ and Al$_{25}$Mo$_{5}$V$_{5}$Ti$_{30}$Cr$_{35}$.  The model predicted low specific mass gains of 0.93 mg/cm$^2$ and 0.79 mg/cm$^2$ for these alloys, respectively. The specific mass gain curves for both 5-component alloys are shown in SI Figure 5. However, the measured oxidation-induced specific mass gains for these compositions were 4.76 mg/cm$^2$ and 5.87 mg/cm$^2$, respectively. These measured values were significantly higher than those of the two best 4-component alloys (Figures ~\ref{fig:microstructure} and ~\ref{fig:interrupted_oxidation}) identified earlier: \ce{Al30Mo5Ti15Cr50} and \ce{Al40Mo5Ti30Cr25}. Consequently, given the observed degree of convergence, the Bayesian optimization process was terminated after six completed batch iterations.

\begin{figure}[H]
    \centering
    \includegraphics[width=\textwidth]{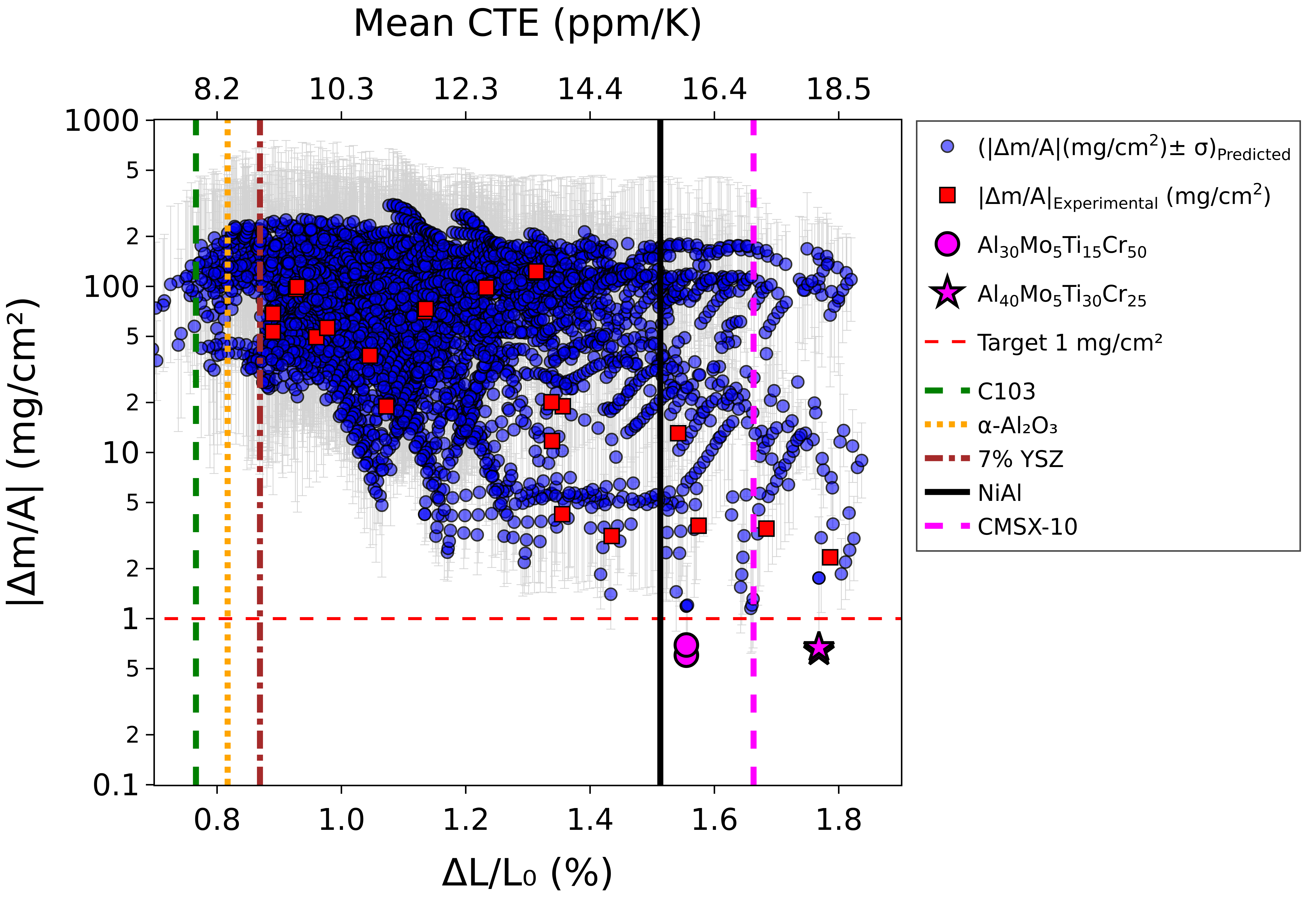}
    \caption{Predicted oxidation resistance (mass gain) versus $\Delta L/L_{o}$\% for RCCAs. Vertical lines represent $\Delta L/L_{o}$\% values of standard ceramic topcoat and metallic bond coat materials. Alloys Al$_{30}$Mo$_5$Ti$_{15}$Cr$_{50}$ and Al$_{40}$Mo$_5$Ti$_{30}$Cr$_{25}$ are highlighted.}
    \label{fig:cte_plot}
\end{figure}

Previous studies have investigated Ti--Al--Cr--X alloys (where X = Si, Hf, Y, Zr, or W) primarily as coatings on $\gamma$-TiAl substrates for high-temperature oxidation protection \cite{Frohlich2008Ti-Al-Cr-XAluminides,Laska2015OxidationCoating,Laska2018OxidationCoating}. These coatings demonstrated improved oxidation resistance through the formation of protective Al$_2$O$_3$ scales and delayed degradation during cyclic exposure at 950--1000\textdegree{}C. While those works focused on the performance of thin films applied to TiAl-based alloys, our study investigates a distinct regime--bulk refractory complex concentrated alloys (RCCAs) containing Al, Ti, and Cr as base elements. In this context, it is particularly interesting to explore the role of Mo addition to the Ti--Al--Cr system, as it may influence not only oxidation behavior but also thermomechanical compatibility in high-temperature environments. Our findings suggest that Mo-containing Ti--Al--Cr--Mo compositions, such as Al$_{30}$Mo$_5$Ti$_{15}$Cr$_{50}$, can form dense $\alpha$-Al$_2$O$_3$ scales while achieving CTE values compatible with common thermal barrier coating (TBC) architectures. This invites further investigation into whether such quaternary or higher-order systems may offer enhanced multifunctional performance beyond the limits of traditional Ti--Al--Cr--X coatings.

The oxidation resistance of our RCCAs, exemplified by Al$_{30}$Mo$_5$Ti$_{15}$Cr$_{50}$ and Al$_{40}$Mo$_5$Ti$_{30}$Cr$_{25}$, aligns with mechanisms observed in Ti--Al--Cr--X coatings. Fröhlich \textit{et al.} \cite{Frohlich2008Ti-Al-Cr-XAluminides} reported that quaternary additions like Y, Zr, and Hf to Ti--Al--Cr coatings on Ti--45Al--8Nb substrates enhanced oxidation resistance at 950--1000\textdegree{}C by stabilizing a cubic B2 phase, which supports sustained Al$_2$O$_3$ scale formation. For instance, the Ti--Al--Cr--Y coating maintained a continuous alumina scale after 1000 cycles at 950\textdegree{}C, with the B2 phase facilitating aluminum diffusion from the $\gamma$-TiAl substrate \cite{Frohlich2010Ti-Al-Cr-TiAl}. Similarly, Lee \textit{et al.} \cite{Lee2002EffectsAlloy} demonstrated that Al--21Ti--23Cr coatings on Ti--48Al exhibited stable oxidation behavior up to 200 hours at 1000\textdegree{}C, enabled by the Ti(Cr,Al)$_2$ phase, which allows alumina formation even at aluminum contents as low as 37--42 at.\%. Our RCCAs, with Al + Cr contents of 45--65 at.\%, exhibit a comparable \textit{Cr effect} in promoting exclusive $\alpha$-Al$_2$O$_3$ formation.

Given the potential of our RCCAs as bond coat materials in thermal barrier coating (TBC) systems, we evaluated their linear thermal expansion ($\Delta L/L_{o}$\%) using Thermo-Calc\textsuperscript{\textregistered} freeze in equilibrium property model calculations at 1000\textdegree{}C---an essential property for thermomechanical compatibility under high-temperature cycling. Although $\Delta L/L_{o}$ was not part of the initial descriptor set, we performed post hoc calculations for all design space alloys to assess their suitability. Figure~\ref{fig:cte_plot} presents a scatter plot of predicted mass gain versus $\Delta L/L_{o}$\%, with vertical dashed lines marking $\Delta L/L_{o}$\% values of reference materials: $\alpha$-Al$_2$O$_3$ (0.81\%), C103 (0.76\%), 7\% YSZ (0.87\%), NiAl (1.53\%), and CMSX-10 (1.66\%).

Our experimentally validated alloy Al$_{30}$Mo$_5$Ti$_{15}$Cr$_{50}$, exhibit $\Delta L/L_{o}$ values of $\sim$1.6\%, positioning them between NiAl and CMSX-10---more compatible with Ni-based superalloys than with $\alpha$-Al$_2$O$_3$ or YSZ. However, Figure~\ref{fig:cte_plot} reveals a critical design trade-off: the predicted trendline indicates that alloys with $\Delta L/L_{o}$ closer to $\alpha$-Al$_2$O$_3$ (0.81\%) can be achieved at the cost of moderately increased mass gain. Specifically, alloys in the 1.0--1.4\% $\Delta L/L_{o}$ range---while exhibiting 2--5$\times$ higher oxidation rates than our optimized compositions---still maintain mass gains well below 10~mg/cm$^2$, remaining within acceptable limits for many TBC applications. This observation suggests a pathway for application-specific optimization: where superior CTE matching with ceramic top coats is prioritized over absolute oxidation minimization, our framework can identify compositions that balance these competing requirements.

The CTE compatibility of our RCCAs enhances their potential as bond coats, offering better adhesion and thermal stability than conventional Ni-based coatings, which can degrade due to phase instability or Al depletion. Unlike Ti-Al-Si or CrN coatings that may form less protective mixed oxides, our RCCAs' ability to generate adherent $\alpha$-Al$_2$O$_3$ layers ensures long-term oxidation resistance. Nevertheless, the observed brittleness---consistent with Ti-Al-Cr alloys---remains a challenge for TBC applications requiring cyclic toughness. Strategies such as microalloying with dopants (e.g., Hf, Y, Si at 0.1--0.5~at.\%) could improve oxide scale adhesion and reduce brittleness, as shown in prior Ti-Al-Cr studies~\cite{Frohlich2010Ti-Al-Cr-TiAl}. The ability of our framework to serendipitously reveal CTE-oxidation trade-offs demonstrates its potential to guide multifunctional property optimization extending beyond single-objective design, enabling targeted alloy selection based on specific TBC system requirements.

The success of our active learning framework, supported by computational tools, lies in its ability to navigate complex design spaces and identify key trends--such as the Al + Cr threshold and the performance advantage of quaternary systems. By selecting Al$_{30}$Mo$_5$Ti$_{15}$Cr$_{50}$ and Al$_{40}$Mo$_5$Ti$_{30}$Cr$_{25}$ as multifunctional alloys combining oxidation resistance, hardness, and CTE compatibility, this study illustrates the transformative potential of data-driven discovery in advancing RCCAs for extreme environments.

\section*{Conclusions}
Through the development of an active learning workflow with verified experimental data continuously improving machine learning models, we conclude:

\begin{enumerate}
\item We developed an active learning framework to identify refractory complex concentrated alloys (RCCAs) that are oxidation resistant upon exposure to air at 1000\textdegree C for 24 h. This framework involved the combined use of Gaussian Process Regression (GPR) with Bayesian Global Optimization (BGO) to efficiently explore the high-dimensional RCCA compositional space. 

\item After only six active learning cycles, involving 5 selected alloys for each cycle, predictions and experimental measurements converged to identify two oxidation-resistant quaternary alloys with nominal compositions of \ce{Al30Mo5Ti15Cr50} and \ce{Al40Mo5Ti30Cr25} (measured compositions of Al$_{26.2}$Mo$_{8.3}$Ti$_{17.2}$Cr$_{48.3}$ and Al$_{42.2}$Mo$_{4.6}$Ti$_{30.3}$Cr$_{22.9}$, respectively). 

\item Both alloys exhibited specific oxidation-induced specific mass gains below 1 mg/cm$^2$ after exposure to air for 24 h at 1000\textdegree C, as confirmed by continuous thermogravimetric analyses during such exposure, and from mass change measurements conducted before and after such exposure. 

\item Raman spectroscopy of the external surfaces, and transmission electron microscopy of ion-milled cross-sections, revealed that both alloys formed continuous, adherent $\alpha$-\ce{Al2O3} scales after 24 h of oxidation at 1000\textdegree C. Such external $\alpha$-\ce{Al2O3} scale formation was consistent with the slow, and decreasing, rates of oxidation observed for these alloys at 1000\textdegree C. 

\item These two alloy compositions also exhibited relatively high values of specific hardness (in excess of 0.12 HV$_{0.5}$m$^3$/kg).

\item Such oxidation resistance and mechanical behavior, coupled with analyses of alloy thermal expansion, suggest that such Al-Ti-Cr-based alloys could be promising candidates as coating materials for other alloy substrates. 

\item This study demonstrates the effectiveness of a scalable, data-driven, active-learning strategy for the rapid identification of oxidation-resistant RCCAs for potential high-temperature applications.
\end{enumerate}
\section*{Methods and Materials}

\subsection*{Alloy Design Space}
The initial design space was constructed to include four-component aluminum (Al) containing alloys. The composition of each element was given in steps of 5 at\%, with the Al content restricted to a range of 5--40 at\% to ensure a focus on Al-containing alloys to promote protective oxide scale formation. This combinatorial approach generated 67,536 distinct alloys across 84 four-component alloy systems.

\begin{figure}[H]
    \centering
    \includegraphics[width=\textwidth]{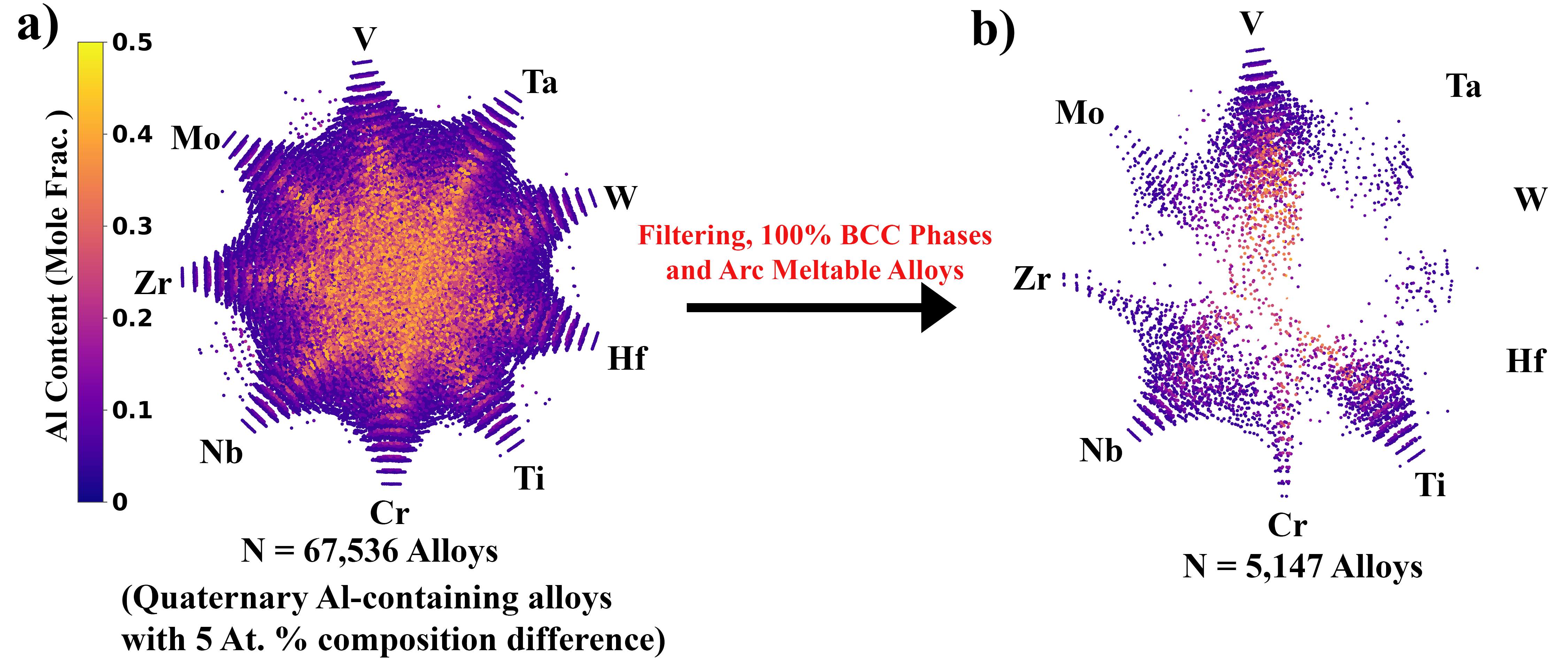}
    \caption{(a) Full RCCA design space mapped using multi-dimensional scaling (MDS). (b) Design space filtered to include only single-phase BCC alloys and to ensure suitability for syntheses by arc melting.}
    \label{fig:design_space}
\end{figure}

This initial design space, illustrated in Figure~\ref{fig:design_space}a, was visualized by reducing the 8-dimensional space of 9 refractory elements (Ti, Zr, Hf, V, Nb, Ta, Cr, Mo, W) to a 2D representation using multi-dimensional scaling, where each point represents an alloy composition, and the color intensity indicated the Al content. Thermodynamic constraints were applied to this initial design space using CALPHAD simulations, following the methodology by Karumuri \textit{et al.} \cite{Austin-Herna2025}. The alloy selection was filtered to ensure suitability for arc melting (i.e., the predicted liquidus temperatures of Al-Cr free alloys needed to be below the boiling points of Al and Cr, which were 2518\textdegree C and 2669\textdegree C, respectively  \cite{Barin1995ReferenceBar}) and for the formation of only body-centered cubic (BCC) phases at 1000\textdegree C to ensure alloy stability during oxidation. Although these additional filters significantly reduced the design space from 67,536 to 5,147 alloys, as depicted in Figure~\ref{fig:design_space}b, these constraints reduced synthesis failures and improved the the correspondence of nominal (targeted) and measured (experimental) alloy compositions.

Additionally, CALPHAD was used to generate physics-based descriptors for the alloys, which served as inputs to the predictive machine-learning model. Single-point equilibrium calculations were conducted for each composition to determine the equilibrium phases, volume fractions, densities, and crystal structures at 1000\textdegree C. Other key thermodynamic properties for each alloy, including the solidus temperature, liquidus temperature, and density, were also calculated. All thermodynamic CALPHAD calculations were conducted using Thermo-Calc\textsuperscript{\textregistered} 2022b software  \cite{Andersson2002Thermo-CalcScience} with the TCHEA4 database, implemented via the TC-Python API (this code is available in the repository \cite{Bejjipurapu2025PyTCPlotter}).

\subsection*{Initial Data}
The initial oxidation dataset used for this study was obtained from Mishra \textit{et al.}  \cite{Mishra2024MassSharing}, in which oxidation mass gain data was compiled for a variety of materials, including pure elements, binary alloys, ternary alloys, and higher-order, high-entropy alloys (HEAs). This dataset contains 408 data points from 145 unique alloy compositions collected from 68 literature sources, with oxidation experiments conducted at various temperatures. All data were standardized and indexed using the FAIR database tool \cite{Saswat2022}. For the present study, we focused on a subset of 81 alloys that underwent oxidation experiments in the temperature range of 900--1000\textdegree C for at least 24 hours. This temperature range and duration aligned with the conditions of our experimental validation trials, in order to ensure relevance to the objectives of this study. Detailed statistics of this dataset, including the distribution of mass gain values and compositional diversity, are provided in Bejjipurapu \textit{et al.} \cite{bejjipurapu2025machinelearningdescriptorspredicting}. 

This initial dataset served as the foundation for training the GPR model and captured the variability in oxidation behavior across a diverse compositional space. Since the mass gain values in the dataset spanned a wide range and exhibited a skewed distribution, and because machine learning models like GPR with BGO typically aim to maximize the target output $y$, the mass gain values were transformed to suit the optimization goal of minimizing the extent of oxidation. We applied the transformation $y = -\ln(\text{mass gain})$, where the negative sign inverted the trend in order to prioritize lower mass gains (since maximizing $-\ln(\text{mass gain})$ corresponded to minimizing mass gain). Moreover, the logarithmic function compressed the wide range of mass gain values into a more manageable scale for modeling. This transformed dataset was used for all subsequent GPR training and BGO optimization steps.

\subsection*{Descriptor Calculations}

We computed a comprehensive set of 34 descriptors representing each alloy in the dataset, enabling the GPR model to correlate alloy composition, structure, and oxidation behavior. These descriptors encompassed fundamental atomic properties, thermodynamic characteristics, and oxidation-specific features. The atomic properties included averages and ranges of elemental attributes such as melting temperature, atomic radii, valence electron concentration (VEC), and Young's moduli, with averages calculated using the rule of mixtures. The thermodynamic descriptors included the entropy of mixing of solutions, solidus and liquidus temperatures, and phase stability metrics derived from CALPHAD simulations. Descriptors specific to oxidation behavior were incorporated to capture the propensity of a given alloy to form protective oxide layers.

To predict the oxide layer structure for each alloy at 1000\textdegree C, a model was employed from Butler \textit{et al.}  \cite{Butler2022OxidationAlloys} based on a database of \textit{ab initio} calculations. This model determines the composition of the oxide scale as a function of the oxygen chemical potential ($\mu_O$) by minimizing the thermodynamic grand potential for oxygen for the system. The grand potential of oxygen ($\overline{\Phi}_G$) is formulated as:

\begin{equation}
\overline{\Phi}_G (\mu_O, N_M, P, T) = \frac{E_0 - \mu_O N_O}{N_M}
\end{equation}

\noindent where $E_0$ is the 0\,K energy of formation from DFT, $\mu_O$ is the chemical potential of oxygen, $N_M$ is the number of metal atoms per formula unit, $N_O$ is the number of oxygen atoms per formula unit, $P$ is the pressure, and $T$ is the temperature. The minimum chemical potential of oxygen for oxide formation was calculated, along with phase fraction weighted Pilling-Bedworth ratio \cite{Xu2000Pilling-bedworthAlloys, Butler2022OxidationAlloys}, and packing efficiencies of the oxide layers extracted directly from the Materials Project database \cite{L.G.WarePyOxidation}. The solidus temperature of the oxide mixture, along with an oxygen solubility ratio (the oxygen solubility in the metal alloy divided by the total mole fraction of oxidizable elements) and the vapor pressure of the oxide mixture at 1000\textdegree C, were calculated using the TCOX 11.0 database in Thermo-Calc\textsuperscript{\textregistered}. One-hot encoding was also used to represent categorical phase data, distinguishing between BCC-only and other phase configurations, as determined by CALPHAD simulations. The definitions and calculation methods for all of the descriptors are provided in the Supplementary Information (SI Table 1). This descriptor set enabled the GPR model to effectively capture the relationship between alloy composition, microstructure, and oxidation mass gain.

\subsection*{Gaussian Process Regression (GPR) Modeling}
Gaussian Process Regression (GPR) was used as a surrogate model to predict the transformed oxidation mass gains ($-\ln(\text{mass gain})$) for untested compositions in the design space, as illustrated in Figure~\ref{fig:workflow} (Gaussian Process Regression segment). The GPR model was trained on the initial dataset, using the calculated descriptors as input features and the transformed mass gain values as the target variable. The GPR model provided both a mean prediction and an uncertainty estimate (predictive variance) for each candidate alloy, which allowed us to quantify model confidence and to guide the selection of compositions for experimental validation. A homoscedastic GPR model was employed throughout the active learning loop, as a single mass gain measurement was typically conducted per condition for each alloy. 
\SK{The algorithm used to implement the homoscedastic GPR model is outlined in Algorithm~\ref{alg:homoscedastic-GPR}.} 
\SK{We optimized the GPR hyperparameters by maximizing the marginal likelihood, a principled approach that balances model fit with complexity. To enhance model interpretability and robustness, we employed Automatic Relevance Determination (ARD), which assigns separate length-scale parameters to each input feature. This allows the model to learn the relative importance of each descriptor by effectively down-weighting irrelevant or less informative features, thereby reducing overfitting and improving generalization across the compositional space.}

\begin{algorithm}[H]
\scriptsize
\caption{\textcolor{blue}{Homoscedastic Gaussian Process Regression (GPR)}}
\label{alg:homoscedastic-GPR}
\textbf{Input:} Training data $\mathcal{D}_{\text{train}} = \{(\mathbf{x}_i, y_i)\}_{i=1}^{n}$, where $\mathbf{x}_i$ are input descriptors and $y_i$ are negative log mass gain values.
\textbf{Assumptions and Notation:}
\vspace{-0.5em}
\begin{enumerate}[label=\alph*)]
    \item Input matrix: $\mathbf{X} = [\mathbf{x}_1^\top, \ldots, \mathbf{x}_n^\top]^\top \in \mathbb{R}^{n \times D}$
    \item Output vector: $\mathbf{y} = [y_1, \ldots, y_n]^\top \in \mathbb{R}^n$
    \item Latent function: $f(\mathbf{x}) \sim \mathcal{GP}(0, k(\mathbf{x}, \mathbf{x}'))$
    \item Noisy observations: $y_i = f(\mathbf{x}_i) + \varepsilon_i$, where $\varepsilon_i \sim \mathcal{N}(0, \beta^2)$
    \item Kernel function: $k(\mathbf{x}, \mathbf{x}') = k_{\text{RBF}}(\mathbf{x}, \mathbf{x}') = \sigma_f^2 \exp\left( -\frac{1}{2} \displaystyle \sum_{d=1}^{D} \frac{(x_d - x_d')^2}{\ell_d^2} \right)$
    with $\sigma_f^2$ denoting the RBF signal variance and $\{\ell_d\}_{d=1}^{D}$ the RBF length-scales.
\end{enumerate}
\textbf{Training Procedure:}
\vspace{-0.5em}
\begin{enumerate}[label=\alph*)]
    \item Compute kernel matrix $\mathbf{K}$ with entries $K_{ij} = k(\mathbf{x}_i, \mathbf{x}_j)$
    \item Prior: $f(\mathbf{X}) \sim \mathcal{N}(0, \mathbf{K})$
    \item Likelihood: $p(\mathbf{y} \mid f(\mathbf{X})) = \mathcal{N}(f(\mathbf{X}), \beta^2 \mathbf{I})$
    \item Marginal likelihood:
    \[
    p(\mathbf{y} \mid \mathbf{X}, \theta) = \mathcal{N}(\mathbf{y} \mid 0, \mathbf{K} + \beta^2 \mathbf{I})
    \]
    \item Log marginal likelihood:
    \[
    \log p(\mathbf{y} \mid \mathbf{X}, \theta) =
    - \frac{1}{2} \mathbf{y}^\top (\mathbf{K} + \beta^2 \mathbf{I})^{-1} \mathbf{y}
    - \frac{1}{2} \log \left| \mathbf{K} + \beta^2 \mathbf{I} \right|
    - \frac{n}{2} \log 2\pi
    \]
    Maximize with respect to hyperparameters $\theta = \{\sigma_f^2, \{\ell_d\}_{d=1}^{D}, \beta^2\}$
\end{enumerate}
\textbf{Prediction at a design space point $\mathbf{x}_*$:}
\vspace{-0.5em}
\begin{enumerate}[label=\alph*)]
    \item Compute:
    \[
    \mathbf{k}_* = k(\mathbf{X}, \mathbf{x}_*) \in \mathbb{R}^n, \quad
    k_{**} = k(\mathbf{x}_*, \mathbf{x}_*) \in \mathbb{R}
    \]
    \item Predictive mean and variance:
    \[
    \begin{aligned}
    m(\mathbf{x}_*) &= \mathbf{k}_*^\top (\mathbf{K} + \beta^2 \mathbf{I})^{-1} \mathbf{y} \\
    \sigma^2(\mathbf{x}_*) &= k_{**} - \mathbf{k}_*^\top (\mathbf{K} + \beta^2 \mathbf{I})^{-1} \mathbf{k}_*
    \end{aligned}
    \]
\end{enumerate}
\textbf{Output:} Predictive mean $m(\mathbf{x}_*)$ and variance $\sigma^2(\mathbf{x}_*)$ at design point $\mathbf{x}_*$, learned hyper-parameters $\theta$.
\end{algorithm}

\subsection*{Bayesian Global Optimization (BGO) with Expected Improvement (EI)}
\begin{figure}[H]
    \centering
    \includegraphics[width=0.5\textwidth]
    {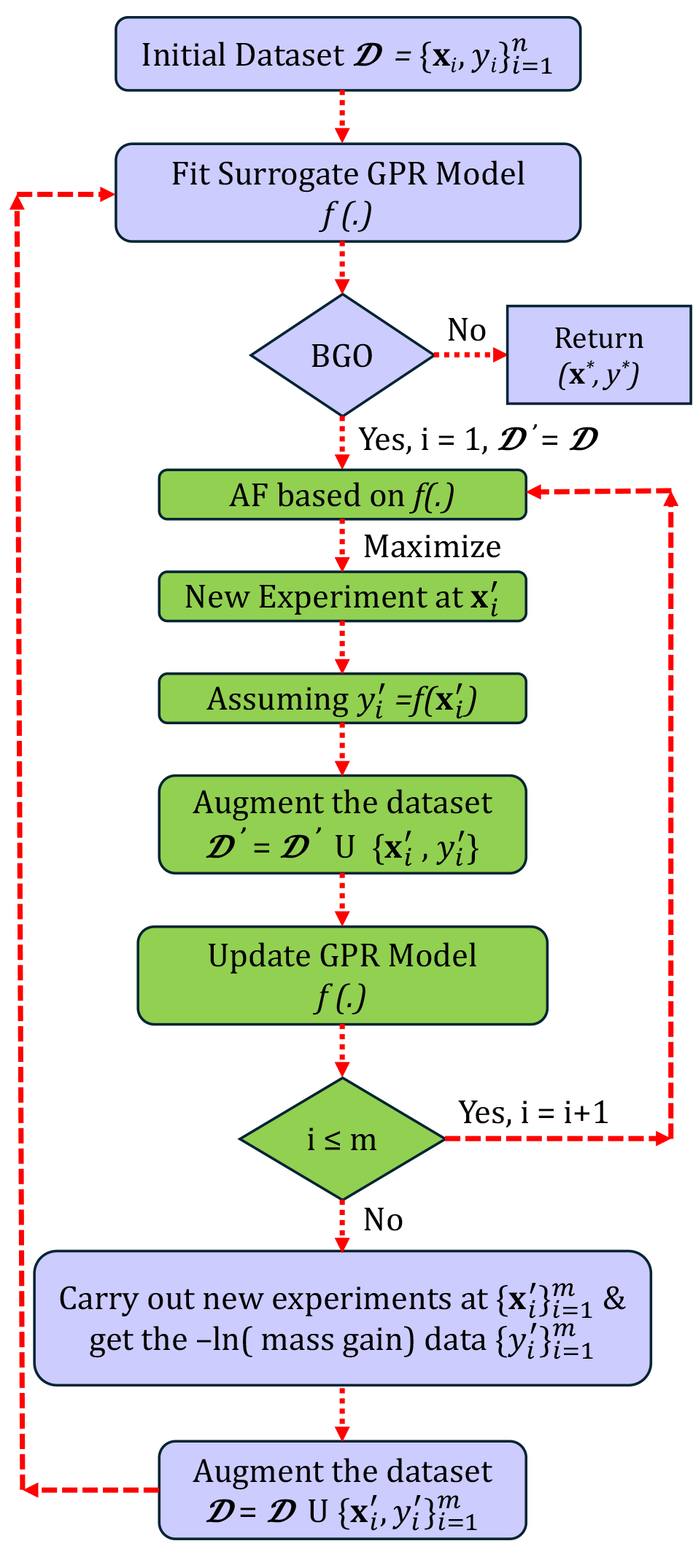}
    \caption{Flowchart of the \SK{Batch Bayesian optimization (Batch - BGO)} approach for iterative alloy selection.}
    \label{fig:bgo_flowchart}
\end{figure}

We implemented BGO to iteratively select compositions for experimental evaluation, as depicted in the flow chart in Figure~\ref{fig:bgo_flowchart}. The optimization problem was formulated as finding the alloy composition \SK{$\mathbf{x}^*$} that maximized the expected value of the transformed target \SK{$\mathbb{E}(y|\mathbf{x})$, where $y = f(\mathbf{x}) + \epsilon$ with $\mathbf{x}$, $y$, $f(\mathbf{x})$, and $\epsilon$ representing the alloy composition, the transformed experimental mass gain,  the transformed predicted mass gain, and the experimental noise, respectively.} Maximizing $y$ in this context corresponds to minimizing the actual mass gain, which was consistent with the goal of identifying oxidation-resistant alloys. 

The Expected Improvement (EI) acquisition was used to balance exploration and exploitation in the active learning loop, as shown in Figure~\ref{fig:workflow} (Batch Bayesian Optimization segment). The EI function leverages the GPR model's predicted mean and standard deviation to prioritize compositions likely to achieve low mass gains (high $-\ln(\text{mass gain})$) while exploring regions of high uncertainty \SK{and exploiting regions of high predictive mean}. \SK{The EI acquisition function is defined as:
\begin{equation}
\label{eqn:EI}
a(\mathbf{x}) = \operatorname{EI}(\mathbf{x}) = (m(\mathbf{x}) - m^*) \ \Phi\left(\frac{m(\mathbf{x}) - m^*}{\sigma(\mathbf{x})}\right)
+ \sigma(\mathbf{x}) \ \phi\left(\frac{m(\mathbf{x}) - m^*}{\sigma(\mathbf{x})}\right)
\end{equation}
where \( m^* \) denotes the best (maximum) value observed so far, while \( \Phi \) and \( \phi \) denote the cumulative distribution function and probability density function of the standard normal distribution, respectively.}

The following steps were followed in each iteration: (1) A GPR surrogate model was constructed using the current dataset, which initially consisted of the transformed mass gain data from Mishra \textit{et al.} \cite{Mishra2024MassSharing, Saswat2022} and was then augmented with new experimental results in subsequent iterations. (2) The EI acquisition function was evaluated for all alloys in the design space to identify promising candidates. (3) A batch of five new alloys was selected that maximized the EI, in order to ensure compositional diversity by iteratively updating the GPR model with synthetic data for each selected alloy within the batch, as outlined in \SK{Figure~\ref{fig:bgo_flowchart} and detailed in Algorithm~\ref{alg:batch-bgo}}.

Specifically, the first alloy was chosen by maximizing EI, the predicted transformed mass gain was added to a copy of the dataset, the GPR model was updated, and the next alloy was then selected using the updated model. This process was repeated until five alloys were selected. This approach prevented the selection of overly-similar alloys and ensured that the batch captured diverse regions of the design space. (4)The selected alloys were synthesized using arc melting and the alloy oxidation behavior was then evaluated through thermogravimetric analysis (TGA) conducted in air at 1000\textdegree C for 24 h, as illustrated in Figure~\ref{fig:workflow} (Experiments segment). (5) The dataset was augmented with the new experimental results, and the measured mass gains were transformed using $-\ln(\text{mass gain})$.
This process was then repeated for the next iteration, as shown in Figure~\ref{fig:workflow} (New Data segment). All synthetic data added during batch generation was removed before starting the next BGO cycle, in order to maintain the integrity of the experimental dataset.

We conducted six such iterations, with each iteration yielding a batch of 5 alloy compositions, and with the design space progressively refined to converge on compositions with notable oxidation resistance. The iterative nature of this workflow, powered by BGO and GPR, enabled us to systematically explore the complex RCCA compositional space, to identify promising alloy candidates, and to validate alloy oxidation resistance through experimental testing, as detailed in the Results and Discussion section.

\subsection*{Fabrication, Testing, and Characterization}
Alloy specimens were fabricated by arc melting of mixtures of elemental constituents, each possessing a purity $\geq$99.7\%: Aluminum (Al),  chromium (Cr), and titanium (Ti) granules, zirconium (Zr) wire form, vanadium (V) foil, hafnium (Hf) wire, niobium (Nb) wire, and tantalum (Ta) wire were obtained from Thermo Scientific Chemicals (Ward Hill, MA USA). Mo wire and W powder were obtained from Midwest Tungsten (Willowbrook, IL USA). The individual elements were weighed using a microbalance (ME36S, Sartorius AG, Weender Landstrasse, Goettingen, Germany) with a resolution of 1 $\mu$g and were then placed on a water-cooled copper hearth within a custom-designed tri-arc melter. The chamber was evacuated to a pressure of 200 mTorr (26.7 Pa) and then backfilled with ultra-high-purity (UHP, 99.999\%) argon. Prior to introduction in the chamber of the tri-arc melter, the UHP argon was passed through a heated-titanium gas purification furnace (model OG-120, Oxy-Gon Industries, Epsom, NH USA)
 
This evacuation and backfilling process was repeated three times to create an inert melting atmosphere. Immediately before arc melting of each alloy, high-purity zirconium granules were arc melted on the copper hearth to act as an additional, localized oxygen getter. Each alloy ingot underwent five sequential melting/solidification cycles, with intermediate flipping of the solidified ingot between each cycle, to promote alloy homogenization. Following arc melting, the alloy ingots were suction cast into a cylindrical copper mold (5 mm diameter $\times$ 20 mm length) using the same type of inert (low oxygen content) atmosphere in the tri-arc melter, but with the water-cooled copper hearth replaced by a high-purity copper mold to facilitate such suction casting. The solidified, suction-cast cylindrical specimens were cross-sectioned into 0.5 mm thick disks using a diamond wafering blade within a low-speed precision saw (TechCut4x, Allied High Tech Products, Cerritos, CA USA). These disks were ground and polished using standard metallographic procedures, culminating in a final polishing step with a 1 $\mu$m diamond suspension to obtain a mirror-like surface finish. The polished samples were then ultrasonically cleaned in ethanol to remove surface contaminants and debris. The nominal compositions of the polished, suction-cast alloys were verified using a scanning electron microscope (Phenom XL SEM, nanoScience Instruments, Phoenix, AZ USA) equipped with energy-dispersive X-ray spectroscopy (EDX). For a given specimen, EDX analysis was conducted over an area of the polished cross-section to obtain the measured alloy composition. 

A preliminary oxidation test was first conducted to determine which of the alloys in a given batch should undergo subsequent thermogravimetric analyses. A polished disk-shaped alloy specimen was placed in a horizontal tube furnace after the furnace had reached 1000 \textdegree C in ambient air. It was held at this temperature for 24 h, after which the sample was removed from the furnace tube and cooled to room temperature. Specimens that exhibited catastrophic oxidation after this preliminary test, as revealed by disk disintegration and/or by complete oxidation (as determined from the mass change) were not examined further. Alloys that retained the disk shape and that were only partially oxidized were then exposed to isothermal oxidation tests using a thermogravimetric analyzer (Jupiter STA 449 F1 TGA, Netzsch Instruments North America, LL, Burlington, MA USA). 
For a given experiment using the TGA, the disk-shaped alloy specimen was placed within a high-purity ($\geq$99.7\% ) alumina crucible (Netzsch Instruments North America, LL, Burlington, MA USA) with the large area faces of the disk in a vertical orientation (i.e., the disk was placed so as to stand on edge in the crucible). The TGA furnace chamber was evacuated and refilled three times with UHP Ar. The specimen was then heated under a constant flow (200 cm$^3$/min) of UHP Ar at 20\textdegree C/min to 1000\textdegree C, and then held at 1000\textdegree C for 15 min to allow for thermal equilibration. The gas environment was then switched to flowing synthetic air (21\% O$_2$, balance N$_2$; 200 cm$^3$/min) for the isothermal oxidation test. The specimen was held in the flowing air for 24 h at 1000\textdegree C during which the specimen mass was continuously evaluated at a data acquisition rate of 10 measurements per sec. After 24 h in flowing air at 1000\textdegree C, the gas was switched back to flowing UHP argon (200 cm$^3$/min) and the furnace was cooled at 20\textdegree C/min to room temperature. Two such oxidation experiments using the TGA were conducted for each alloy specimen that had been selected using the preliminary oxidation test discussed above.
After oxidation, the surface oxide present on each specimen was characterized using a Raman spectrometer (InVia, Renishaw Inc., West Dundee, IL, USA) equipped with a 633 nm He-Ne laser as the excitation source. To ensure uniformity in spectral acquisition and comparability of the resulting Raman spectra, each Raman analysis was conducted by exposing a laser spot of 5 $\mu$m diameter at the center of the disk-shaped specimen for 5 sec. Ten such Raman spectra were collected at the same location for each specimen to improve the signal-to-noise ratio.

To further investigate the oxidation behavior of the most promising alloys, the nominal alloy compositions with the lowest mass gains, Al$_{30}$Mo$_{5}$Ti$_{15}$Cr$_{50}$ and Al$_{40}$Mo$_{5}$Ti$_{30}$Cr$_{25}$, were selected for additional heat treatment and time-resolved oxidation studies. To allow for further chemical homogeneity (relative to the as-cast dendritic microstructures), the suction-cast specimens were annealed for 10 h at 1300\textdegree C under a vacuum of 1 $\times$ 10$^{-6}$ (T-M Vacuum VH-HV Flipper furnace, T-M Vacuum Products, Inc., NJ, USA). These homogenized samples underwent intermittent oxidation tests in ambient air using a horizontal tube furnace. After heating the furnace to 1000\textdegree C, the samples were rapidly inserted into the isothermal zone and then removed at specified intervals of 1, 4, and 24 h. Both alloys were also subjected to thermogravimetric oxidation tests using the previously described conditions. The nanoscale structure and composition of the oxide scale formed on the homogenized Al$_{30}$Mo$_{5}$Ti$_{15}$Cr$_{50}$ alloy after oxidation for 1 h and 24 h at 1000°C was evaluated by Transmission Electron Microscopy (TEM). TEM lamellae were prepared using a focused-ion-beam milling system (Helios G4 UX DualBeam, Thermo Scientific, Waltham, MA USA), operated at 30 kV with ion beam currents ranging from 30 pA to 15 nA. Site-specific cross-sections were extracted and transferred onto copper TEM grids using a lift-out system (Omniprobe 200, Oxford Instruments, Abingdon, UK). TEM analyses (Talos 200X, Thermo Scientific) were conducted using an accelerating voltage of 200 keV. EDX analyses were performed in scanning TEM (STEM) mode, and high-resolution imaging was conducted using a high-angle annular dark-field (HAADF) detector. HAADF-STEM images were acquired at a camera length of 205 mm, using a collection angle range of 28-172 mrad. Electron diffraction patterns were compared to simulated patterns, using the CrystalMaker and SingleCrystal software, for phase identification. 
\section*{Acronyms}
\begin{tabular}{@{}ll@{}}
\textbf{RCCA} & Refractory Complex Concentrated Alloy \\
\textbf{HEA} & High Entropy Alloy \\
\textbf{MPEA} & Multi-Principal Element Alloy \\
\textbf{ML} & Machine Learning \\
\textbf{AI} & Artificial Intelligence \\
\textbf{GPR} & Gaussian Process Regression \\
\textbf{BGO} & Bayesian Global Optimization \\
\textbf{EI} & Expected Improvement \\
\textbf{BCC} & Body-Centered Cubic \\
\textbf{CALPHAD} & Calculation of Phase Diagrams \\
\textbf{DFT} & Density Functional Theory \\
\textbf{SEM} & Scanning Electron Microscope \\
\textbf{BSE} & Backscattered Electron \\
\textbf{EDX} & Energy-Dispersive X-ray Spectroscopy \\
\textbf{STEM} & Scanning Transmission Electron Microscopy \\
\textbf{TEM} & Transmission Electron Microscopy \\
\textbf{SAED} & Selected Area Electron Diffraction \\
\textbf{HAADF} & High-Angle Annular Dark Field \\
\textbf{TGA} & Thermogravimetric Analysis \\
\textbf{CTE} & Coefficient of Thermal Expansion \\
\textbf{$\Delta L/L_{o}$} & Linear thermal expansion \\
\textbf{VEC} & Valence Electron Concentration \\
\textbf{PBR} & Pilling-Bedworth Ratio \\
\textbf{UHP} & Ultra-High Purity \\
\textbf{TBC} & Thermal Barrier Coating \\
\textbf{YSZ} & Yttria-Stabilized Zirconia \\
\textbf{HV} & Vickers Hardness \\
\textbf{RBF} & Radial Basis Function \\
\textbf{ARD} & Automatic Relevance Determination \\
\textbf{k\textsubscript{p}} & Parabolic Rate Constant \\
\end{tabular}

\section*{Supplementary Information}

The supplementary material includes the pseudocode for the active learning strategy (SI Algorithm 1), a complete list of all 34 descriptors used in the GPR model (SI Table 1), and a comprehensive table of nominal compositions, measured compositions, and mass gain data for all tested alloys (SI Table 2). Supplementary figures include the exploration vs. exploitation plots for each active learning batch (SI Figure 1), the relationship between Raman oxide peak intensity and mass gain (SI Figure 2), SEM-EDX cross-sections of the oxide scales on the \ce{Al40Mo5Ti30Cr25} (SI Figure 3) and \ce{Al30Mo5Ti15Cr50} (SI Figure 4) alloys, and the TGA results for the two 5-component alloys (SI Figure 5).

\section*{Author contributions}
\textbf{Akhil Bejjipurapu:} Writing – original draft; Data curation; Formal analysis; Investigation; Methodology; Software; Validation; Visualization. \\
\textbf{Sharmila Karumuri:} Writing – review \& editing; Methodology. \\
\textbf{Joseph C. Flanagan:} Writing – review \& editing; Investigation. \\
\textbf{Victoria Tucker:} Investigation (TEM Analysis). \\
\textbf{Ilias Bilionis:} Conceptualization; Funding acquisition; Project administration; Resources; Supervision; Writing – review \& editing. \\
\textbf{Alejandro Strachan:} Conceptualization; Funding acquisition; Project administration; Resources; Supervision; Writing – review \& editing. \\
\textbf{Kenneth H. Sandhage:} Conceptualization; Funding acquisition; Project administration; Resources; Supervision; Writing – review \& editing. \\
\textbf{Michael S. Titus:} Conceptualization; Funding acquisition; Project administration; Resources; Supervision; Writing – review \& editing.

\section*{Conflicts of interest}
There are no conflicts to declare.

\section*{Data availability}

The data and computer codes that support the findings of this study are available in the cited references. Any additional data are available from the corresponding author upon reasonable request.
\section*{Acknowledgements}

We acknowledge support from the U.S. National Science Foundation, DMREF program, under Contract No. 1922316-DMR.

\newpage
\begingroup
\sloppy
\printbibliography
\endgroup


\newpage
\appendix
\section{Supplementary Information} 
\setcounter{page}{1} 
\renewcommand{\thepage}{A\arabic{page}} 
\setcounter{algorithm}{0} 
\renewcommand{\thealgorithm}{SI \arabic{algorithm}} 
\setcounter{figure}{0} 
\renewcommand{\thefigure}{SI\arabic{figure}} 
\renewcommand{\figurename}{SI Figure} 
\setcounter{table}{0} 
\renewcommand{\thetable}{SI\arabic{table}} 
\renewcommand{\tablename}{SI Table} 


\begin{algorithm}[H]
\scriptsize
\caption{\textcolor{blue}{Our active learning strategy (Batch - BGO).}}
\label{alg:batch-bgo}
\begin{algorithmic}[1]
\Require Literature dataset $\mathcal{D}_{\text{lit}} = \{(\mathbf{x}_i, y_i)\}_{i=1}^{n_{\text{lit}}}$, 
 Candidate design set $\mathbf{X}_{\text{design}} = [\mathbf{x}_1^\top, \ldots, \mathbf{x}_{n_{\text{design}}}^\top]^\top \in \mathbb{R}^{n_{\text{design}} \times D}$, Batch index $b$, batch size $m$, acquisition function $a(\mathbf{x}) = \text{EI}(\mathbf{x})$.
 \If{$b = 1$} 
 \State $\mathcal{D}_{\text{train}} \gets \mathcal{D}_{\text{lit}}$ \Comment{Initialization with literature data}
\EndIf
\If{$b \geq 1$}  
 \State $\mathcal{D}'_{\text{train}} \gets \mathcal{D}_{\text{train}}$ 
    \State Fit a homoscedastic GPR model $f(\cdot)$ on $\mathcal{D}'_{\text{train}}$ using Algorithm~\ref{alg:homoscedastic-GPR} ; extract hyper-parameters $\theta = \left\{ \sigma_f^2, \{\ell_d\}_{d=1}^{D}, \beta^2 \right\}$
    \For{$i = 1$ to $m$}  \Comment{Suggest an alloy batch}
        \State Predict posterior mean and variance $m(\mathbf{x}), \sigma^2(\mathbf{x})$ for all $\mathbf{x} \in \mathbf{X}_{\text{design}}$
        \State Compute acquisition $a(\mathbf{x})$ using $m(\mathbf{x})$, $\sigma(\mathbf{x})$, and $m^* = \max_{\mathbf{x}_i \in \mathcal{D}'_{\text{train}}} m(\mathbf{x}_i)$
        \State Select $\mathbf{x}_i^b = \arg\max_{\mathbf{x} \in \mathbf{X}_{\text{design}}} a(\mathbf{x})$
        \State Generate virtual observation: $\tilde{y}_i^b = m(\mathbf{x}_i^b) + \beta \epsilon, \ \epsilon \sim \mathcal{N}(0, 1)$
        \State Augment: $\mathcal{D}'_{\text{train}} \gets \mathcal{D}'_{\text{train}} \cup \{(\mathbf{x}_i^b, \tilde{y}_i^b)\}$
        \State Update homoscedastic GP model $f(\cdot)$ using $\mathcal{D}'_{\text{train}}$
    \EndFor
    \State \Return Suggested batch $\{\mathbf{x}_i^b\}_{i=1}^{m}$.
    \State Perform physical experiments at $\{\mathbf{x}_i^b\}_{i=1}^{m}$ to obtain true responses $\{y_i^b\}_{i=1}^{m}$.  
    \State Augment: $\mathcal{D}_{\text{train}} \gets \mathcal{D}_{\text{train}} \cup \{(\mathbf{x}_i^b, y_i^b)\}_{i=1}^{m}$ \Comment{Store this}
\EndIf
\end{algorithmic}
\end{algorithm}

\vspace{1cm} 
\author{}  
\date{}   
\maketitle
\renewcommand{\arraystretch}{2.0}
\begin{longtable}{@{}c>{\raggedright\arraybackslash}p{2cm}>{\centering\arraybackslash}m{5cm}>{\raggedright\arraybackslash}p{6cm}@{}}
\caption*{\textbf{SI Table 1:} Full list of 34 descriptors used in the Gaussian Process Regression (GPR) model. Formulations correspond to mathematical expressions used for feature computation, where applicable.}
\label{tab:supp_SI_descriptors} \\
\toprule
\textbf{Index} & \textbf{Notation} & \textbf{Formulation} & \textbf{Descriptor Description} \\
\midrule
\endfirsthead
\toprule
\textbf{Index} & \textbf{Notation} & \textbf{Formulation} & \textbf{Descriptor Description} \\
\midrule
\endhead
\bottomrule
\endfoot
1 & $T_m$ & $\sum_{i=1}^{n} c_i T_{m,i}$ & Average melting temperature \\
2 & $V_{\text{misfit}}$ & $\sum_{i=1}^{n} (\bar{V} - V_i)^2$ & Atomic volume misfit \\
3 & $\overline{R}$ & $\sum_{i=1}^{n} c_i r_{\text{at},i}$ & Average atomic radius \\
4 & $\delta$ & $\sqrt{\sum_{i=1}^{n} c_i\left(1 - \left(\frac{r_i}{\bar{r}}\right)\right)^2} \times 100$ & Asymmetry of atomic radii \\
5 & VEC & $\sum_{i=1}^{n} c_i \cdot \text{VEC}_i$ & Average valence electron concentration \\
6 & $\Delta S_{\text{mix}}$ & $-R \sum_{i=1}^{n} c_i \ln c_i$ & Entropy of mixing \\
7 & $|Y|$ & $\sqrt{\sum_{i=1}^{n} c_i \left(1 - \left(\frac{Y_i}{\bar{Y}}\right)\right)^2} \times 100$ & Asymmetry of Young's moduli \\
8 & $\Delta Y$ & $Y_{\text{max}} - Y_{\text{min}}$ & Range of Young's moduli \\
9 & $\Delta \rho$ & $\rho_{\text{max}} - \rho_{\text{min}}$ & Range of density \\
10 & $\Delta T_m$ & $T_{m,\text{max}} - T_{m,\text{min}}$ & Range of melting temperature \\
11 & $\Delta r_{\text{at}}$ & $r_{\text{at},\text{max}} - r_{\text{at},\text{min}}$ & Range of atomic radii \\
12 & $\Delta K$ & $K_{\text{max}} - K_{\text{min}}$ & Range of bulk moduli \\
13 & $\Delta$VEC & $\text{VEC}_{\text{max}} - \text{VEC}_{\text{min}}$ & Range of VEC \\
14 & $\rho$ & -- & Density of the alloy in kg/m$^3$ \\
15 & Solidus & -- & Solidus temperature of the alloy \\
16 & Liquidus & -- & Liquidus temperature of the alloy \\
17 & $\mu_O$ & -- & Minimum chemical potential of oxygen at which there is an oxide with a volume fraction $>$ 0.30 (\textbf{\textit{First 30-Vol\% Layer}}). \\
18 & PBR & -- & Weighted Pilling--Bedworth Ratio of the \textbf{\textit{First 30-Vol\% Layer}} \\
19 & Protective Oxide Layer & 0 / 1 & Presence of Al$_{2}$O$_{3}$, Cr$_{2}$O$_{3}$, or SiO$_{2}$ in \textbf{\textit{First 30-Vol\% Layer}} \\
20--22 & Packing Efficiency$_{\text{max,min,avg}}$ & -- & Max, min, and average packing efficiency among oxides in the \textbf{\textit{First 30-Vol\% Layer}} \\
23 & Oxide Volume Fraction & -- & Volume fraction of oxide in the \textbf{\textit{First 30-Vol\% Layer}} \\
24 & Solidus of Oxide Mixture & -- & Solidus temperature of the oxide mixture in the \textbf{\textit{First 30-Vol\% Layer}} \\
25 & Solidus of All Oxide Layers & 0 / 1 & If any oxide layer has solidus $<$ 1000°C \\
26 & Oxygen Solubility Ratio & 
$\frac{N_O}{N_B} = \frac{\text{Oxygen solubility of the bulk alloy}}{\sum\limits_{i \neq O} composition\ of\ element_i}$ & 
Ratio of oxygen solubility in the alloy (N$_O$) to the sum of the mole fraction of the oxidizing elements in the bulk alloy (N$_B$) \\
27 & Vapor Pressure & -- & Weighted vapor pressure of oxide mixture at 1000°C \\
28 & pO$_2$ & -- & Partial pressure of oxygen during oxidation \\
29--34 & BCC,\ BCC+FCC,\ BCC+Sec.\newline
FCC,\ FCC+Sec.,\ Other Phase & [1,0,0,...,0],[0,1,0,...,0] etc & 
One-hot encoding for phases \\
\end{longtable}
\newpage
\renewcommand{\arraystretch}{1.8}
\setlength{\tabcolsep}{3pt} 
\begin{longtable}{|c|>{\centering\arraybackslash}p{3.4cm}|>{\centering\arraybackslash}p{4.4cm}|>{\centering\arraybackslash}p{1.7cm}|>{\centering\arraybackslash}p{1.7cm}|>{\centering\arraybackslash}p{1.7cm}|}
\caption*{\textbf{SI Table 2.} Nominal and measured compositions, along with predicted, measured, and repeated mass gain values for 30 refractory complex concentrated alloys (RCCAs). For alloys that volatilized completely or reacted with the Al$_{2}$O$_{3}$ crucible, repeated tests were not conducted. Alloy rows are color-coded by campaign for clarity.}

\label{tab:mass_gain_colored_subscript} \\
\hline
\rowcolor{gray!30}
Campaign & Nominal Composition (At. $\%$) & Measured Composition (At.$\%$) & Predicted Mass Gain (mg/cm$^2$) & Measured Mass Gain (mg/cm$^2$) & Repeated Mass Gain (mg/cm$^2$) \\
\hline
\endfirsthead
\hline
\rowcolor{gray!30}
Campaign & Nominal Composition (At. $\%$) & Measured Composition (At.$\%$) & Predicted Mass Gain (mg/cm$^2$) & Measured Mass Gain (mg/cm$^2$) & Repeated Mass Gain (mg/cm$^2$) \\
\hline
\endhead
\hline
\endfoot
\rowcolor{blue!10} AL - B1 & Al$_5$Nb$_{65}$Zr$_5$Ti$_{25}$ & Al$_{2.2}$Nb$_{68.3}$Zr$_{4.7}$Ti$_{24.7}$ & 3.3 & 69.5 & - \\
\rowcolor{blue!10} AL - B1 & Al$_5$Mo$_{70}$Ti$_{20}$W$_5$ & Al$_{3.4}$Mo$_{72.2}$Ti$_{19.0}$W$_{5.4}$ & 14.3 & 56.5 & - \\
\rowcolor{blue!10} AL - B1 & Al$_{15}$V$_{35}$W$_{15}$Cr$_{35}$ & Al$_{17.8}$V$_{32.8}$W$_{13.2}$Cr$_{36.2}$ & 20.4 & 98.4 & - \\
\rowcolor{blue!10} AL - B1 & Al$_{15}$V$_{35}$Ti$_{20}$Cr$_{30}$ & Al$_{13.3}$V$_{37.5}$Ti$_{21.2}$Cr$_{28.1}$ & 21.6 & 121.2 & 120.5 \\
\rowcolor{blue!10} AL - B1 & Al$_5$Hf$_5$Mo$_{80}$Ta$_{10}$ & Al$_{4.0}$Hf$_{4.8}$Mo$_{78.4}$Ta$_{12.8}$ & 21.8 & 30.5 & - \\
\rowcolor{red!10} AL - B2 & Al$_{20}$Mo$_5$Ta$_5$Ti$_{70}$ & Al$_{24.3}$Mo$_{4.4}$Ta$_{4.8}$Ti$_{66.5}$ & 1.5 & 4.3 & 4.9 \\
\rowcolor{red!10} AL - B2 & Al$_5$Nb$_5$Ta$_{35}$Ti$_{55}$ & Al$_{6.7}$Nb$_{4.6}$Ta$_{34.7}$Ti$_{54.0}$ & 28.8 & 56.4 & 56.9 \\
\rowcolor{red!10} AL - B2 & Al$_5$Hf$_5$Ti$_{85}$Cr$_5$ & Al$_{9.0}$Hf$_{6.3}$Ti$_{79.2}$Cr$_{5.5}$ & 30.8 & 19.0 & 19.1 \\
\rowcolor{red!10} AL - B2 & Al$_{25}$Mo$_{10}$Ta$_5$Ti$_{60}$ & Al$_{26.6}$Ti$_{59.5}$Mo$_{8.3}$Ta$_{5.5}$ & 2.2 & 3.1 & 3.2 \\
\rowcolor{red!10} AL - B2 & Al$_5$V$_5$Nb$_{85}$Zr$_5$ & Al$_{7.8}$V$_{4.6}$Nb$_{83.6}$Zr$_{4.0}$ & 35.7 & 96.4 & 97.0 \\
\rowcolor{green!10} AL - B3 & Al$_{15}$V$_{35}$Ti$_{20}$Cr$_{30}$ & Al$_{16.3}$V$_{32.9}$Ti$_{21.2}$Cr$_{29.6}$ & 2.3 & 121.2 & 120.8 \\
\rowcolor{green!10} AL - B3 & Al$_{35}$Nb$_5$Ti$_{50}$Cr$_{10}$ & Al$_{33.1}$Nb$_{4.9}$Ti$_{46.1}$Cr$_{17.9}$ & 2.4 & 3.5 & 3.5 \\
\rowcolor{green!10} AL - B3 & Al$_{15}$Mo$_{35}$Ti$_{35}$Cr$_{15}$ & Al$_{12.2}$Mo$_{33.1}$Ti$_{37.1}$Cr$_{17.6}$ & 2.4 & 70.8 & 71.0 \\
\rowcolor{green!10} AL - B3 & Al$_{20}$Mo$_{10}$Nb$_5$Ti$_{65}$ & Al$_{17.3}$Mo$_{11.1}$Nb$_{6.5}$Ti$_{65.1}$ & 3.1 & 11.7 & 11.8 \\
\rowcolor{green!10} AL - B3 & Al$_{30}$Ta$_5$Ti$_{50}$Cr$_{15}$ & Al$_{33.9}$Ta$_{3.2}$Ti$_{52.9}$Cr$_{10.0}$ & 4.7 & 3.6 & 3.7 \\
\rowcolor{orange!10} AL - B4 & Al$_{40}$Mo$_5$Ti$_{30}$Cr$_{25}$ & Al$_{36.3}$Mo$_{5.7}$Ti$_{35.7}$Cr$_{22.3}$ & 0.8 & 0.6 & 0.7 \\
\rowcolor{orange!10} AL - B4 & Al$_{30}$Mo$_5$Ti$_{15}$Cr$_{50}$ & Al$_{27.9}$Mo$_{6.3}$Ti$_{16.1}$Cr$_{49.6}$ & 3.2 & 0.6 & 0.6 \\
\rowcolor{orange!10} AL - B4 & Al$_{15}$Mo$_{35}$Ti$_{35}$Cr$_{15}$ & Al$_{17.4}$Mo$_{33.1}$Ti$_{39.1}$Cr$_{10.4}$ & 1.2 & 70.3 & 70.5 \\
\rowcolor{orange!10} AL - B4 & Al$_{25}$Mo$_{25}$Nb$_{15}$Ti$_{35}$ & Al$_{21.9}$Mo$_{24.3}$Ti$_{37.0}$Nb$_{16.7}$ & 0.9 & 19.0 & 19.1 \\
\rowcolor{orange!10} AL - B4 & Al$_{15}$Mo$_{35}$Ti$_{40}$Cr$_{10}$ & Al$_{18.1}$Mo$_{36.1}$Ti$_{43.1}$Cr$_{2.6}$ & 1.0 & 73.2 & 73.5 \\
\rowcolor{purple!10} AL - B5 & Al$_{15}$V$_{35}$Ti$_{20}$Cr$_{30}$ & Al$_{19.2}$Ti$_{19.3}$V$_{33.2}$Cr$_{28.3}$ & 1.1 & 123.0 & 120.2 \\
\rowcolor{purple!10} AL - B5 & Al$_5$Nb$_{10}$Zr$_{15}$Ti$_{70}$ & Al$_{7.0}$Ti$_{68.6}$Zr$_{12.2}$Nb$_{12.2}$ & 9.0 & 38.3 & 38.5 \\
\rowcolor{purple!10} AL - B5 & Al$_{40}$V$_{25}$W$_5$Cr$_{30}$ & Al$_{48.0}$V$_{21.2}$Cr$_{26.6}$W$_{4.2}$ & 0.6 & 2.3 & 2.4 \\
\rowcolor{purple!10} AL - B5 & Al$_5$Nb$_5$Ta$_{65}$Ti$_{25}$ & Al$_{3.0}$Ti$_{27.9}$Nb$_{7.8}$Ta$_{61.3}$ & 17.2 & 53.3 & 53.5 \\
\rowcolor{purple!10} AL - B5 & Al$_{30}$Mo$_5$Ti$_{15}$Cr$_{50}$ & Al$_{26.2}$Ti$_{17.2}$Cr$_{48.3}$Mo$_{8.3}$ & 2.4 & 0.9 & 0.9 \\
\rowcolor{cyan!10} AL - B6 & Al$_{30}$Mo$_{10}$Nb$_5$Ti$_{55}$ & Al$_{31.1}$Ti$_{56.5}$Nb$_{4.1}$Mo$_{8.3}$ & 5.1 & 13.1 & 13.2 \\
\rowcolor{cyan!10} AL - B6 & Al$_{20}$Nb$_5$Ta$_{15}$Ti$_{60}$ & Al$_{18.3}$Ti$_{61.1}$Nb$_{3.3}$Ta$_{17.3}$ & 5.7 & 20.1 & 20.2 \\
\rowcolor{cyan!10} AL - B6 & Al$_5$V$_{45}$Ta$_{25}$W$_{25}$ & Al$_{4.9}$V$_{48.2}$Ta$_{26.2}$W$_{20.7}$ & 21.7 & 99.4 & 99.5 \\
\rowcolor{cyan!10} AL - B6 & Al$_{40}$Mo$_5$Ti$_{30}$Cr$_{25}$ & Al$_{42.2}$Ti$_{30.3}$Cr$_{22.9}$Mo$_{4.7}$ & 1.8 & 0.7 & 0.7 \\
\rowcolor{cyan!10} AL - B6 & Al$_5$Nb$_{85}$Ta$_5$W$_5$ & Al$_{6.9}$Nb$_{81.3}$Ta$_{4.7}$W$_{7.1}$ & 25.4 & 68.7 & 69.0 \\
\hline
\end{longtable}
\setlength{\tabcolsep}{6pt} 
\newpage
\begin{figure}[h]
\centering
\includegraphics[width=\textwidth]{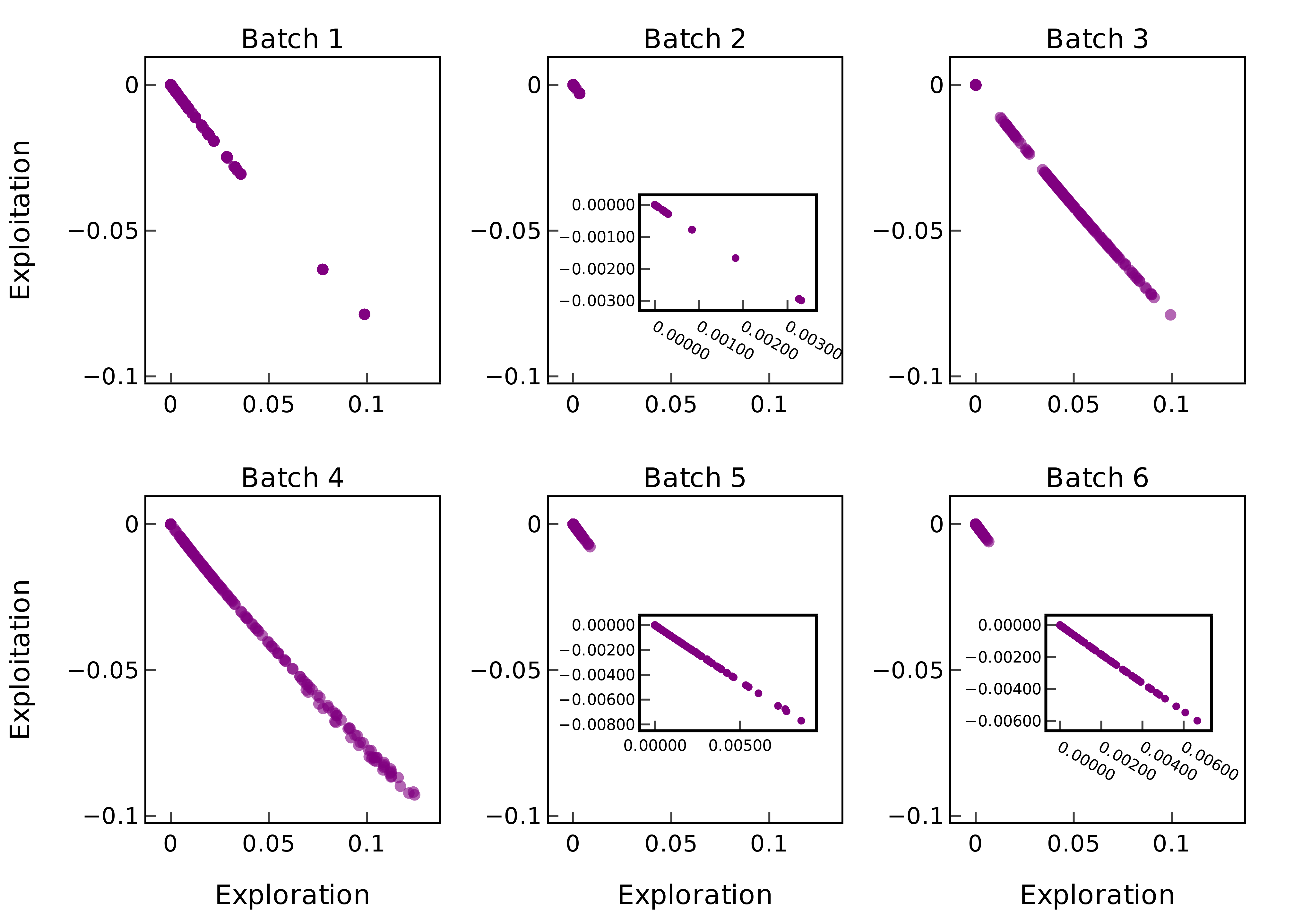}
\caption*{\textbf{SI Figure 1.} Scatter plots of exploration versus exploitation values for all alloys in Batches 1–6 during batch Bayesian optimization. Insets in Batches 2, 5, and 6 highlight regions with low exploration and exploitation values. As active learning progresses, a clear trade-off emerges—early batches (e.g., Batch 1) favor broad exploration, while later batches (e.g., Batch 4–6) increasingly exploit the surrogate model's predictions.}
\end{figure}
\begin{figure}[h]
\centering
\includegraphics[width=\textwidth]{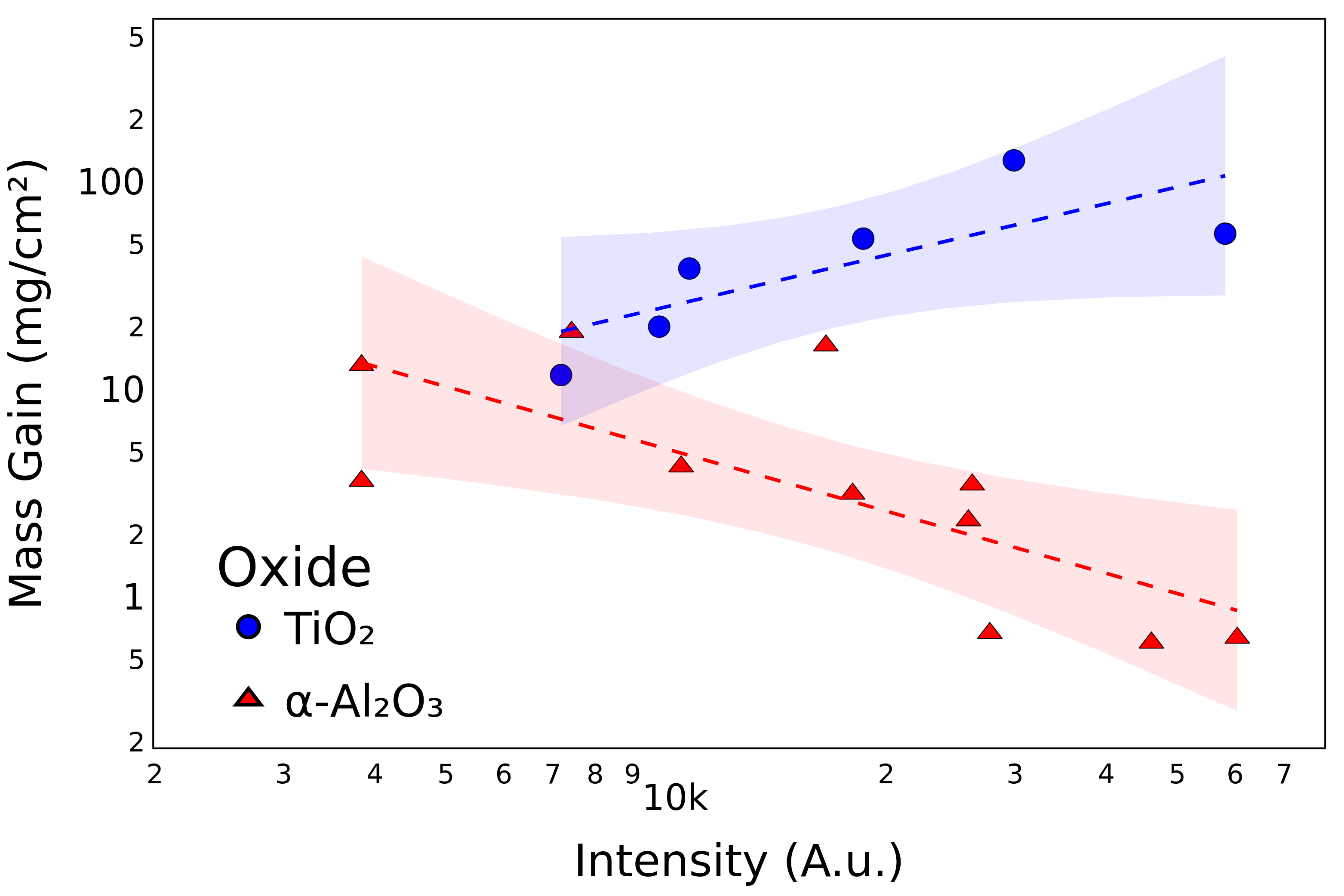}
\caption*{\textbf{SI Figure 2.} Relationship between Raman oxide peak intensity and 24-hour mass gain for selected alloys. Shaded regions represent 95\% confidence intervals for the regression fits.}
\end{figure}

\begin{figure}[h]
\centering
\includegraphics[width=\textwidth]{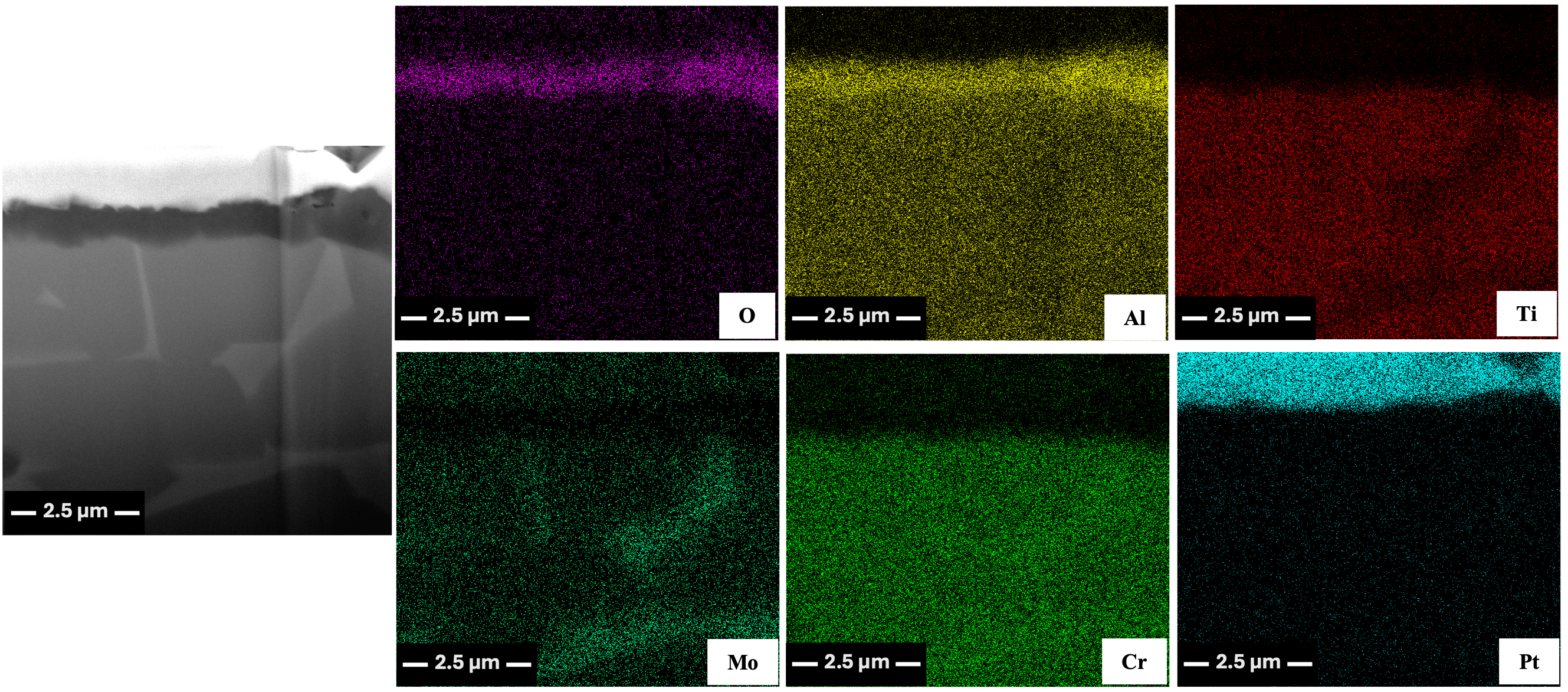}
\caption*{\textbf{SI Figure 3.} SEM image and EDX elemental maps of a FIB-milled cross-section through the oxide scale and underlying substrate for the Al$_{40}$Mo$_{5}$Ti$_{30}$Cr$_{25}$ alloy oxidized in TGA at 1000\,\textdegree{}C for 24 hours.}
\end{figure}
\begin{figure}[h]
\centering
\includegraphics[width=\textwidth]{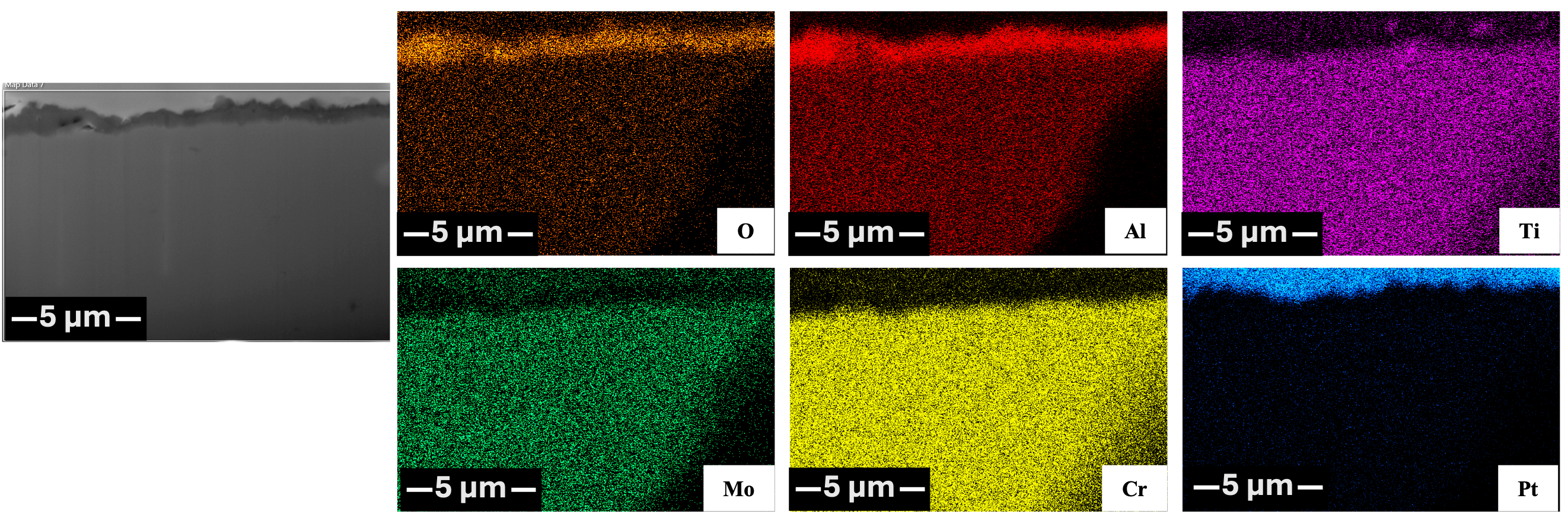}
\caption*{\textbf{SI Figure 4.} SEM image and EDX elemental maps of a FIB-milled cross-section through the oxide layer and underlying substrate for the Al$_{30}$Mo$_{5}$Ti$_{15}$Cr$_{50}$ alloy oxidized in TGA at 1000\,\textdegree{}C for 24 hours.}
\end{figure}
\begin{figure}[h]
\centering
\includegraphics[width=\textwidth]{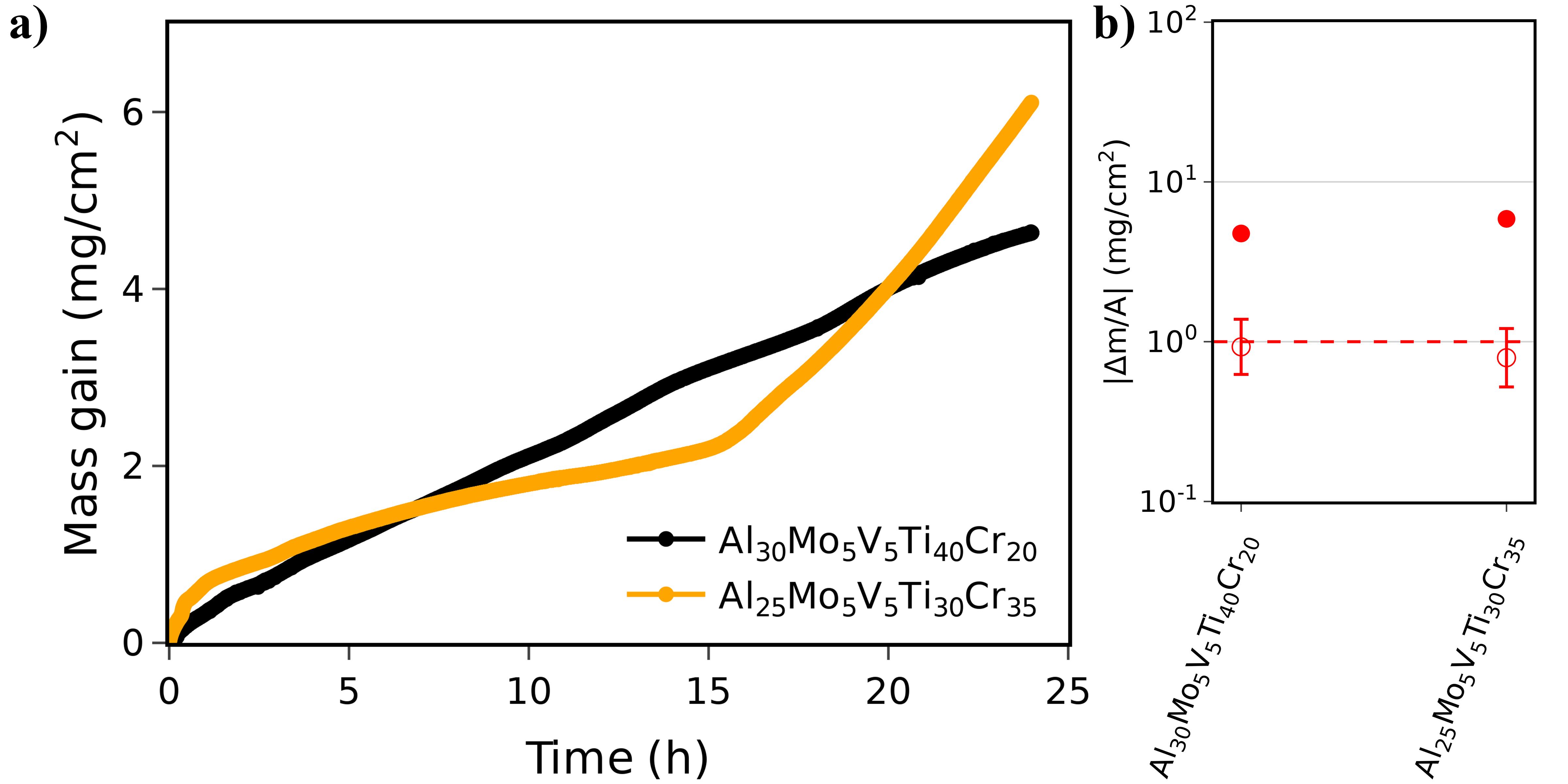}

\caption*{\textbf{SI Figure 5.}(\textbf{a}) Thermogravimetric analysis (TGA) plots showing mass gain as a function of time for Al$_{30}$Mo$_{5}$V$_{5}$Ti$_{40}$Cr$_{20}$ (black) and Al$_{25}$Mo$_{5}$V$_{5}$Ti$_{30}$Cr$_{35}$ (yellow), tested in laboratory air at 1000\,\textdegree{}C for 24 hours. 
(\textbf{b}) Comparison of predicted and measured mass gain values. Open red circles represent values predicted by the Gaussian Process Regression (GPR) model with their standard deviations, and filled red circles represent experimentally measured values.}
\end{figure}

\end{document}